\documentclass[aps, prl, 10pt, twocolumn, superscriptaddress, floatfix]{revtex4-1}

\usepackage{amsmath}
\usepackage{upgreek}
\usepackage{graphicx}
\usepackage{lmodern}
\usepackage[margin=0.65in]{geometry}
\usepackage{silence}
\usepackage{hyperref}
\usepackage{soul, color} 
\hypersetup
{
colorlinks=true,
citecolor=blue
}
\def\ll#1{}

\begin{document}

\title{A Recurrent Ising Machine in a Photonic Integrated Circuit}

\author{Mihika Prabhu$^{\bullet}$}  \email{mihika@mit.edu}
 \affiliation{MIT Research Lab of Electronics, 50 Vassar St, Cambridge, MA 02139, USA} 
 \author{Charles Roques-Carmes$^{\bullet}$} \email{chrc@mit.edu}
 \affiliation{MIT Research Lab of Electronics, 50 Vassar St, Cambridge, MA 02139, USA} 
\author{Yichen Shen$^{\bullet}$}\email{ycshen@mit.edu}
 \affiliation{Lightelligence, 268 Summer Street, Boston, MA 02210}
\author{Nicholas Harris}
 \affiliation{Lightmatter, 61 Chatham St 5th floor, Boston, MA 02109, USA}
\author{Li Jing} 
 \affiliation{MIT Department of Physics, 77 Massachusetts Avenue, Cambridge, MA 02139, USA} 
\author{Jacques Carolan} 
 \affiliation{MIT Research Lab of Electronics, 50 Vassar St, Cambridge, MA 02139, USA} 
\author{Ryan Hamerly} 
 \affiliation{MIT Research Lab of Electronics, 50 Vassar St, Cambridge, MA 02139, USA} 
\author{Tom Baehr-Jones} 
 \affiliation{Elenion Technologies, LLC, 171 Madison Ave 1100, New York, NY 10016, USA} 
\author{Michael Hochberg} 
 \affiliation{Elenion Technologies, LLC, 171 Madison Ave 1100, New York, NY 10016, USA} 
\author{Vladimir \v{C}eperi\'{c}} 
 \affiliation{MIT Department of Physics, 77 Massachusetts Avenue, Cambridge, MA 02139, USA} 
\author{John D. Joannopoulos} 
 \affiliation{MIT Department of Physics, 77 Massachusetts Avenue, Cambridge, MA 02139, USA} 
 \affiliation{Institute for Soldier Nanotechnologies, 500 Technology Square, Cambridge, MA 02139, USA \\ 
 $^{\bullet}$denotes equal contribution.}
 \author{Dirk R. Englund}
 \affiliation{MIT Research Lab of Electronics, 50 Vassar St, Cambridge, MA 02139, USA} 
\author{Marin Solja\v{c}i\'{c}}
 \affiliation{MIT Research Lab of Electronics, 50 Vassar St, Cambridge, MA 02139, USA} 
  \affiliation{MIT Department of Physics, 77 Massachusetts Avenue, Cambridge, MA 02139, USA} 

\begin{abstract}
Conventional computing architectures have no known efficient algorithms for combinatorial optimization tasks, which are encountered in fundamental areas and real-world practical problems including logistics, social networks, and cryptography. Physical machines have recently been proposed and implemented as an alternative to conventional exact and heuristic solvers for the Ising problem, one such optimization task that requires finding the ground state spin configuration of an arbitrary Ising graph. However, these physical approaches usually suffer from decreased ground state convergence probability or universality for high edge-density graphs or arbitrary graph weights, respectively. We experimentally demonstrate a proof-of-principle integrated nanophotonic recurrent Ising sampler (INPRIS) capable of converging to the ground state of various 4-spin graphs with high probability. The INPRIS exploits experimental physical noise as a resource to speed up the ground state search. By injecting additional extrinsic noise during the algorithm iterations, the INPRIS explores larger regions of the phase space, thus allowing one to probe noise-dependent physical observables. Since the recurrent photonic transformation that our machine imparts is a fixed function of the graph problem, and could thus be implemented with optoelectronic architectures that enable GHz clock rates (such as passive or non-volatile photonic circuits that do not require reprogramming at each iteration), our work paves a way for orders-of-magnitude speedups in exploring the solution space of combinatorially hard problems.
\end{abstract}

\maketitle

Combinatorial optimization is critical for a broad array of tasks, including artificial intelligence, bioinformatics, cryptography, scheduling, trajectory planning, and circuit design \cite{Landau2009, Hromkovic2013, Aarts1988}. However, combinatorial problems typically fall into the nondeterministic-polynomial hard (NP-hard) problem class, becoming computationally intractable at large scale for traditional algorithms. This challenge motivates the search for alternatives to conventional (von Neumann) computing architectures that can efficiently solve such problems. The Ising problem, which consists of finding the ground state spin configuration of a quadratic Hamiltonian defined by a symmetric matrix $\mathbf{K}$ and spins of unit amplitude $\sigma_{1\leq i\leq N} \in \{-1,1\}^N$,
\begin{equation}
	H^{(K)}(\vec{\sigma})=-\frac{1}{2}\sum_{1\leq i,j\leq N} \sigma_{i}K_{ij}\sigma_{j},
\label{Eq:Ham}
\end{equation}
has garnered significant attention as many other combinatorial problems can be polynomially reduced to an Ising problem \cite{Karp1972, Lucas2014}. Therefore, any technique for finding the ground state of arbitrary Ising problems, which is an NP-hard computational task, may extend to a wide range of other computationally intensive optimization problems. 


There is currently no known efficient classical algorithm to find the exact ground state of an arbitrary Ising graph, so heuristic and meta-heuristic algorithms are often implemented as a means of quickly obtaining approximate solutions \cite{Glover2006}. Various physical systems have been proposed as Ising machines, as the evolution of many natural systems (ferromagnets \cite{Ising1925}, lipid membranes \cite{Honerkamp-Smith2009}, random photonic networks \cite{Ghofraniha2015}, etc.) can be described by Hamiltonians similar to the one in Eq.~(\ref{Eq:Ham}). 
Parallel machines provide additional advantages by reducing the correlation between consecutive samples and preventing premature trapping in local minima by applying many local modifications simultaneously \cite{Macready1996} and can also efficiently explore the phase space with many independent searches running in parallel \cite{Tsukamoto2017, Landau2009, Earl2005}. 


This observation motivates the development of (re)configurable parallel analog machines, such as programmable nanophotonic processors \cite{Carolan2015, Shen2017a, Harris2017, Harris2018a,Roques-Carmes2018}. The computational speedup with these machines is still polynomial (since analog machines also suffer from the P = NP paradigm \cite{Vergis1986}). However, when operating at a fast clock rate (GHz), afforded by a photonic implementation, these optical architectures could enable orders-of-magnitude speedups against conventional solvers \cite{Roques-Carmes2018}.

\begin{figure*}[t]
\centering
\includegraphics[width = \textwidth]{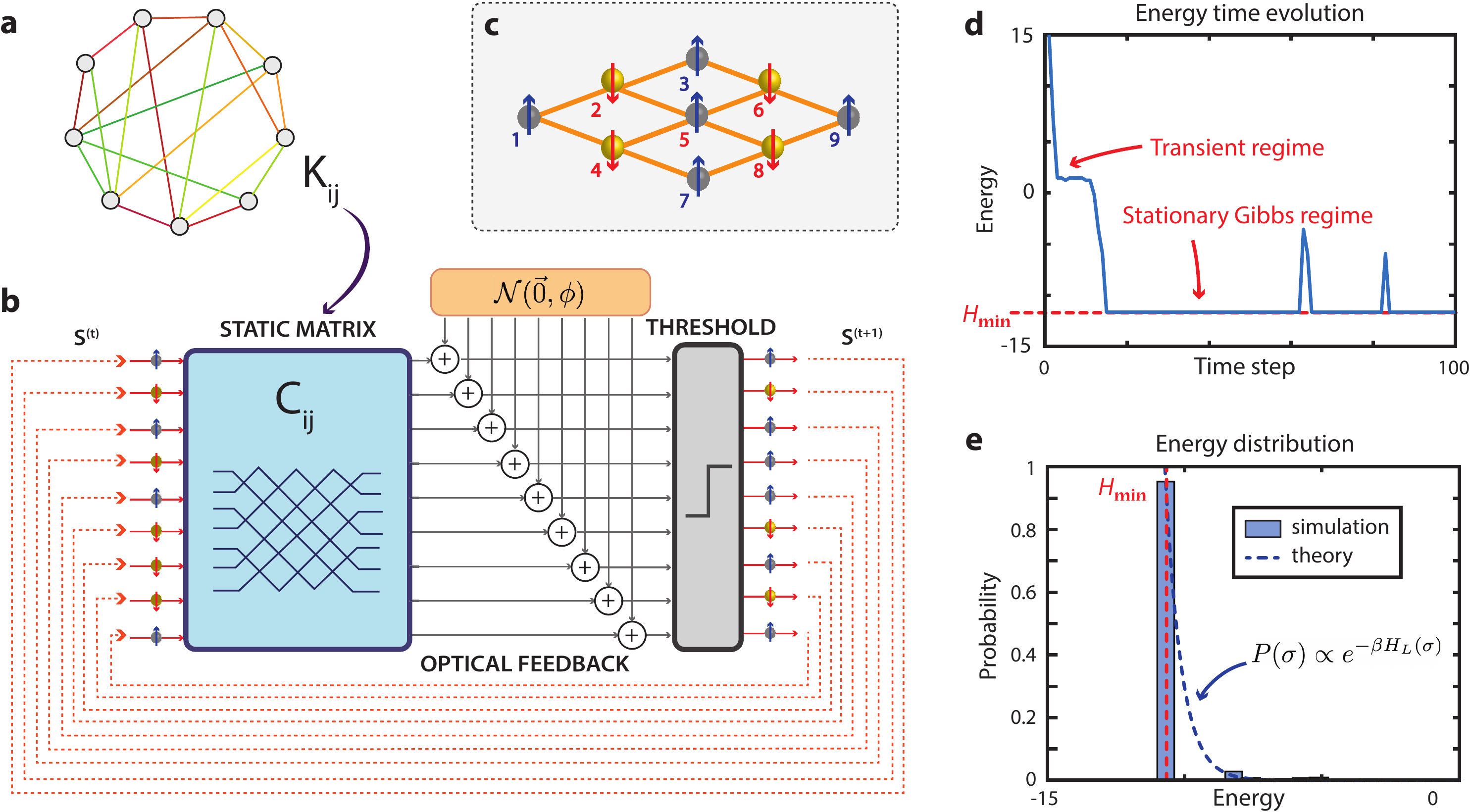}
\caption{\textbf{Photonic Recurrent Ising Sampler.} \textbf{(a)} An arbitrary Ising graph described by a coupling matrix $\mathbf{K}$ is processed to produce the fixed transformation for the recurrent sampler, shown in \textbf{(b)}.  \textbf{(b)} A single algorithm iteration of the photonic recurrent Ising sampler. An in-phase optical signal encoding a spin state is fed to an optical matrix multiplication unit encoding $\mathbf{C}=2\sqrt{\mathbf{K}}$, where $\mathbf{K}$ is the coupling matrix of the desired Ising graph. The output signal is noisy, with a distribution that is Gaussian with  standard deviation $\phi$, and goes through an analog nonlinear unit before being fed back to the chip input. Considering as an example a 9-spin 2D antiferromagnetic graph, with coupling and ground state shown in \textbf{(c)}, the simulated energy evolution as a function of time is shown in \textbf{(d)}. \textbf{(e)} Simulated energy distribution of the optical output, which converges in probability to the Gibbs distribution of the associated Ising problem (Eq \ref{eq:gibbs}), for which the ground state \textbf{(c)} is exponentially more likely than higher-energy states at low temperatures.  }
\label{fig:gen_concept}
\end{figure*}

\begin{figure*}[t]
\centering
\includegraphics[width = \textwidth]{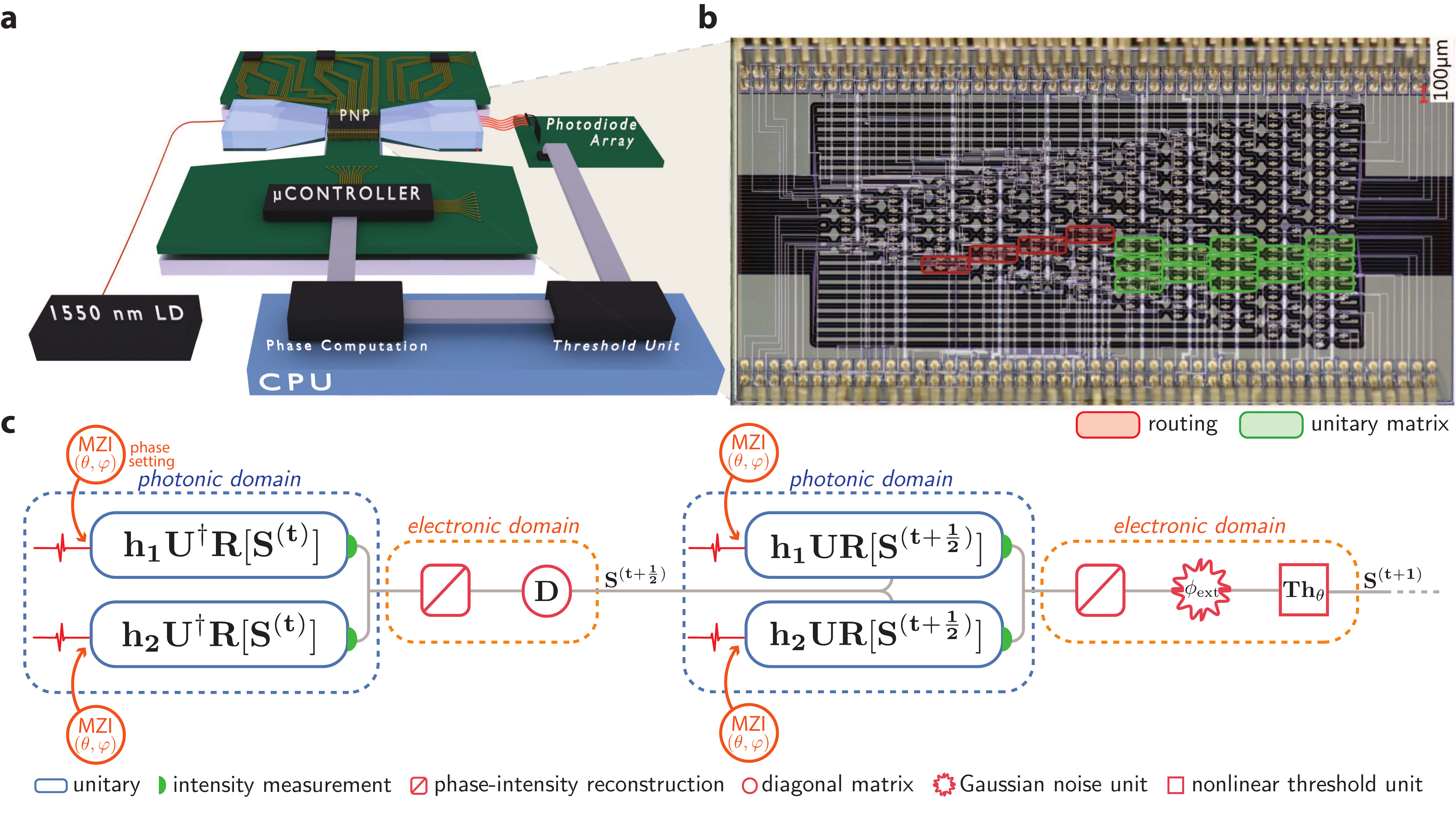}
\caption{\textbf{Experimental realization of Integrated Nanophotonic Coherent Ising Sampler.} \textbf{(a)} A 1550nm laser diode is coupled to a single input port of a silicon-on-insulator programmable nanophotonic processor (PNP). Output mode intensities are measured with an InGaAs photodiode array, then processed in the electronic domain to determine PNP phase settings for the next iteration. \textbf{(b)} The PNP is comprised of 88 Mach-Zehnder interferometers with 176 individually-controlled thermal phase shifters \cite{Harris2017, Harris2018a} and encodes a circuit consisting of input routing (red) and a U(5) unitary matrix (green). \textbf{(c)} Each algorithm step, shown conceptually in Figure \ref{fig:gen_concept}(b), requires four passes through the PNP chip. Each use of the chip performs a unitary matrix product of a state-preparation rotation matrix $\mathbf{R}$, the desired unitary ($\mathbf{U}$ or $\mathbf{U^{\dagger}}$), and one of two homodyne detection matrices ($\mathbf{h_{1}}$ or $\mathbf{h_{2}}$). Phase-intensity reconstruction, a diagonal matrix multiplication, Gaussian noise addition, and a nonlinear threshold unit are applied in the electronic domain. }
\label{fig:setup}
\end{figure*}

\begin{figure*}
\centering
\includegraphics[width = \textwidth]{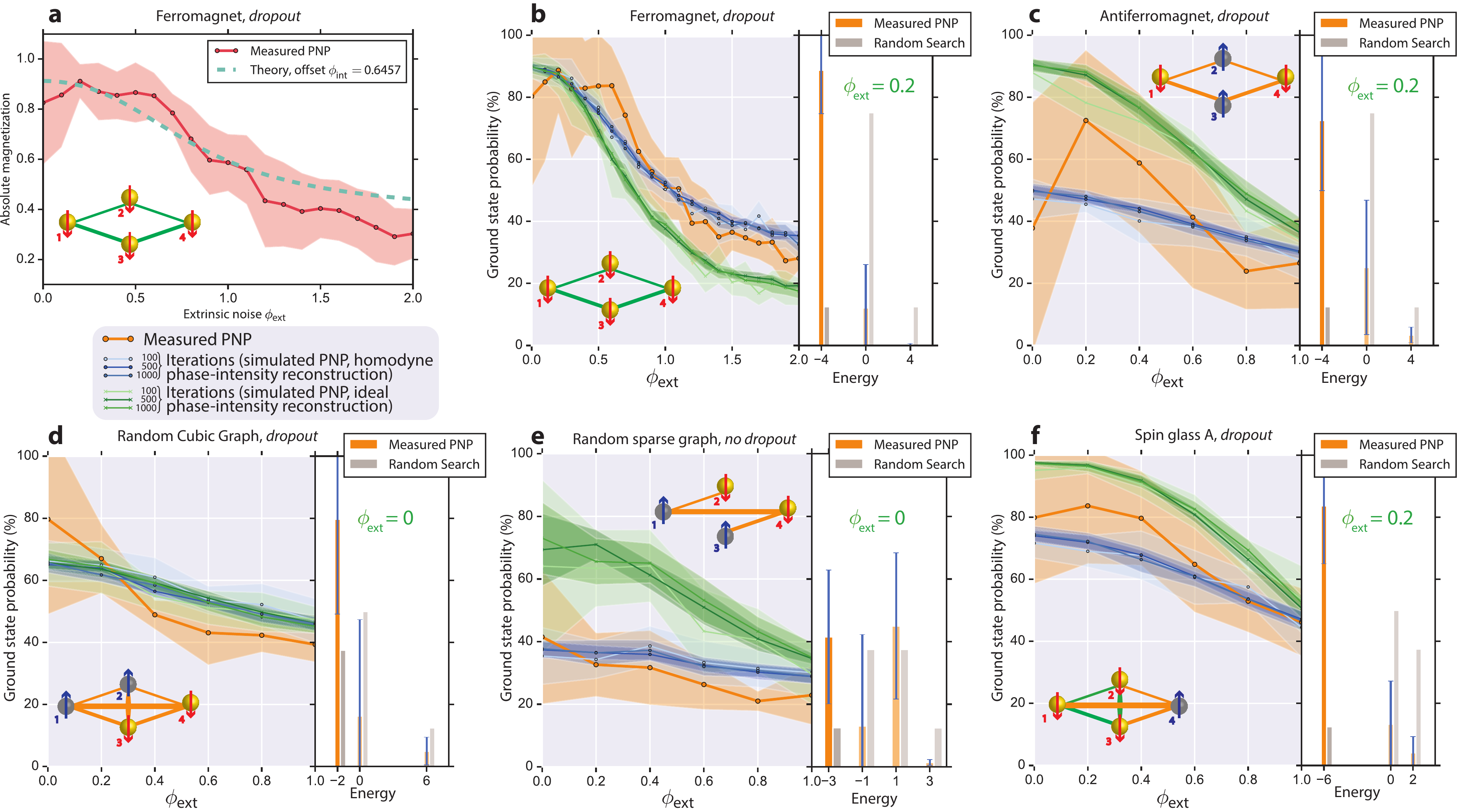}
\caption{\textbf{Evaluating physical observables and finding ground states with the Integrated Nanophotonic Coherent Ising Sampler.} \textbf{(a)} Experimental evaluation of the magnetization of a two-dimensional ferromagnet (red dots). The fit with theory allows us to evaluate intrinsic noise on the PNP $\phi_\text{int} = 0.6457$. \textbf{(b-f)} \textbf{Left:} Ground state probability as a function of extrinsic noise $\phi_\text{ext}$ (orange), compared to simulated PNP with homodyne (blue) and ideal (green) phase-intensity reconstruction. \textbf{Right:} Energy histograms for value of $\phi_\text{ext}$ showing best performance and comparison with random search. \textbf{Inset:} Schematic representation of the Ising model being modeled by the PNP. Green (resp. orange) lines between nodes represent +1 (resp. -1) couplings.}
\label{fig:results}
\end{figure*}


A number of photonic Ising machines have been developed in the past few years, based on various physical systems including degenerate optical parametric oscillators (OPO) \cite{Wang2013, McMahon2016}, fiber networks \cite{Wu2014, Vazquez2018}, and spatial multiplexing \cite{Pierangeli2019}. Some of these (classical) Ising machines even demonstrated performance superior to that of certain quantum annealers on dense graphs \cite{Hamerly2018}. Experimental OPO demonstrations \cite{McMahon2016, Hamerly2018} and recent theoretical proposals \cite{Leleu2019} have shown great potential for finding optimal cuts of MAX-CUT problems with large numbers of spins. However, all-optical OPO machines face scaling limitations due to dispersion and decoherence of the time-division-multiplexed pulses in increasingly large resonators \cite{Inagaki2016}. Additionally, some hybrid systems implement couplings via matrix multiplication in a field-programmable gate array (FPGA), and thus facing similar scaling issues to those found in traditional computing for large-scale multiplication \cite{McMahon2016}. 


Here, we experimentally demonstrate a photonic recurrent Ising sampler for probabilistically finding the ground state of an arbitrary Ising problem. Using a programmable silicon-on-insulator nanophotonic processor \cite{Little1974,Hopfield1982, Peretto1984, Roques-Carmes2018,Harris2017,Shen2017a}, we leverage schemes for decomposing any unitary matrix into a mesh of linear optical components \cite{Clements2016,Reck1994} to enable sampling of arbitrary Ising graphs. Furthermore, if an $N\times N$ matrix is encoded in the optical domain, one can perform a matrix-vector multiplication with an $N$-dimensional vector, represented by $N$ optical fields, using approximately $N$ operations instead of the $N^{2}$ operations required by traditional electronic matrix-vector multiplication. Recent work on parallel photonic circuits for optical neural networks has demonstrated this capability for $O(N)$ vector-matrix multiplication with an array of tunable Mach-Zehnder interferometers (MZIs) \cite{Shen2017a}.  

The proof-of-principle integrated nanophotonic recurrent Ising sampler (INPRIS) is based on a recently proposed parallel photonic recurrent network designed to find the ground state energies of Ising problems \cite{Roques-Carmes2018}. The conceptual structure of a single algorithm step of the $N$-spin photonic recurrent network is shown in Figure \ref{fig:gen_concept}(b). At time step $t$, the spin state vector $\vec{S}^{(t)} \in \{ 0, 1 \}^{N}$ (equivalent to $\sigma_{i}^{(t)} = 2S_{i}^{(t)} - 1 \in \{ -1, 1 \}$) is encoded in the amplitudes of $N$ coherent optical signals at the input. During each algorithm step, the optically encoded spin state vector propagates through an array of optical components that encodes an arbitrarily reconfigurable $N\times N$ optical matrix \cite{Clements2016, Shen2017a, Harris2017}, $\mathbf{C}$, that is fixed and dependent on the problem-specific Ising coupling matrix, $\mathbf{K}$. The output of the matrix multiplication is noisy, being perturbed by a Gaussian noise source with standard deviation $\phi$. This Gaussian perturbation to the signal could be realized with electrical or optical modulation of the refractive index, or via quantum detection noise \cite{Hamerly2018b}. The output of the noisy matrix multiplication is then fed to an optoelectronic threshold operation, $f_{\vec{\theta}_\text{th}}$, where $\vec{\theta}_\text{th}$ is a vector of threshold values, that converts the continuous output vector back into a binary spin state, $\vec{S}^{(t+1)} \in \{ 0, 1 \}^{N}$. In combination, the sampler calculates the input spins for the subsequent algorithm step using the following expression:
\begin{eqnarray}
\vec{S}^{(t+1)} = f_{\vec{\theta}_\text{th}} \left( \mathcal{N} \left( \mathbf{C}\vec{S}^{(t)}, \phi \right) \right),\label{eq:alg_step}
\end{eqnarray}
where $\mathcal{N}(\mu, \phi)$ is the normal distribution with mean, $\mu$, and standard deviation, $\phi$.

After many algorithm steps, the output spin state probability distribution converges to an Ising Gibbs distribution (independent of the initial spin state) \cite{Roques-Carmes2018}:
\begin{eqnarray}
\lim_{t \to \infty}P \left( \vec{\sigma}^{(t)} \right) &\propto & \exp \left[ -\beta H_L \left( \vec{\sigma}^{(t)} \right) \right], \label{eq:gibbs}\\
H_L \left(\vec{\sigma} \right) & = & - \frac{1}{\beta} \sum_i \log \cosh \left( \beta \sum_j J_{ij} \sigma_j \right)\\
& \approx & \beta H^{(\mathbf{J}^T \mathbf{J})}(\vec{\sigma}),\label{eq:ham_little}
\end{eqnarray}
where $\mathbf{C} = 2 \mathbf{J}$ is the matrix implemented by the optical circuit, and $\beta = 1/(k\phi)$ ($k$ being a constant that depends only on the noise distribution present in the system). The effective Hamiltonian of the system $H_L$ can be approximated by an Ising Hamiltonian with weights $\mathbf{J}^2$ for small $\beta$ (corresponding to large noise, $\phi$) and symmetric $\mathbf{J}$. Therefore, to probe a symmetric Ising coupling matrix, $\mathbf{K}=\mathbf{UDU}^{\dagger}$, with $\mathbf{UDU}^\dagger$ being the unitary decomposition of $\mathbf{K}$, we program the optical matrix to implement the square root of $\mathbf{K}$. The square root motivates us to express the optically applied diagonal matrix in the form $\mathbf{D}_\alpha = \text{Re} \left( \sqrt{\mathbf{D} + \alpha \mathbf{\Delta}}\right)$. The term $\alpha \mathbf{\Delta}$ is a diagonal offset matrix with $\mathbf{\Delta} > 0$, and $\alpha$ is a scalar dropout parameter that allows us to tune the dimensionality of the ground state search by selectively dropping out lower negative eigenvalues (the choice of parameters is described in \cite{Roques-Carmes2018}). In the following, we will refer to $\alpha = 1$ as \textit{no dropout} (none of the eigenvalues are dropped out, as $D + \alpha \Delta$ is positive semidefinite) and $\alpha = 0$ as \textit{dropout} (the negative eigenvalues of $D + \alpha \Delta$ are set to zero by taking the real part). In our algorithm, the eigenvalue dropout parameter is a fixed transformation on the coupling matrix, unlike dropout encountered in machine learning, which is a dynamic technique to prevent overfitting when training neural networks \cite{Srivastava}.

To illustrate an example ground state search of the parallel photonic network architecture, we simulate a two-dimensional antiferromagnet with periodic boundary conditions and identical nearest-neighbor couplings, $K_{ij}=-1$, shown in Figure \ref{fig:gen_concept}(c). After several algorithm steps, the optical output converges to a probability distribution that matches the Gibbs distribution with energy described in Eq. (\ref{eq:gibbs}). Thus, the ground state with energy $H_\text{min}$ (every spin having opposite direction to its nearest neighbours, shown in Figure \ref{fig:gen_concept}(c)) is most likely, which can be readily observed in the simulated optical output (Figure \ref{fig:gen_concept}(d)). The expected Gibbs-like energy distribution is also readily observable on the simulated energy output distribution (Figure \ref{fig:gen_concept}(e)).


The experimental INPRIS demonstration presented in this work finds the ground state of various 4-spin Ising problems with arbitrary graph couplings, using noise as a resource to enable convergence to the ground state. The optical matrix multiplication update step, $\mathbf{C}\vec{S}^{(t)}$ (from Eq. (\ref{eq:alg_step})), is broken down into three stages, motivated by the eigenvalue decomposition of the Ising weight matrix $\mathbf{K}$. The two unitary matrix multiplication steps, $\mathbf{U}^{\dagger}$ and $\mathbf{U}$, are implemented in a programmable nanophotonic processor (PNP) \cite{Harris2017, Harris2018a} comprised of a unitary mesh of tunable MZIs (Figure \ref{fig:setup}(b)). The diagonal matrix  $\mathbf{D}_\alpha$ is currently performed on an external CPU but could also be implemented optically with an electro-optic attenuator, a modulator, or a single MZI \cite{Shen2017a, Cheng2014, Bao2011a}. 

Our experimental setup (Figure \ref{fig:setup}) consists of a 1550 nm laser feeding one port of the PNP, then routed (red MZIs in Figure \ref{fig:setup}(b)) to a $5\times 5$ unitary processor comprised of 13 individually thermally-tunable MZIs \cite{Harris2014} (green MZIs in Figure \ref{fig:setup}(b)), controlled by a microcontroller. The output intensities of each spatial waveguide in the PNP are measured by an InGaAs photodiode array and subjected to the set of nonlinear transformations outlined in Eq. (\ref{eq:alg_step}) and Figure \ref{fig:setup} (c) (phase-intensity reconstruction, addition of extrinsic Gaussian noise, and nonlinear threshold unit), currently implemented in an external CPU. The output of these electronic transformations then determines the PNP phase setting at the next algorithm step, and so forth. 

To maximize the number of spins that can be handled by a single PNP, we broke down the algorithm step shown in Figure \ref{fig:gen_concept}(b) into multiple runs on the PNP, such that each run requires only one matrix to be encoded on the chip (shown in Figure \ref{fig:setup}(c)). During each run, the PNP implements a unitary matrix of the form $\mathbf{h}_i \mathbf{V R}[ S^{(\tau)}]$  ($i \in {1, 2}$, Figure \ref{fig:setup}(c)), where $\mathbf{R}[S^{(\tau)}]$ rotates the PNP input $(1,0,0,0,0)^T$ into vector $S^{(\tau)}$, $\tau$ is the integer (resp. half integer) number of algorithm steps, and $S^{(\tau)}$ is the current spin state (resp. an intermediate result after multiplication by $\mathbf{DU}^\dagger$). $\mathbf{V}$ corresponds to one of the two unitary matrices involved in the eigenvalue decomposition of $\mathbf{K}$ ($\mathbf{V} \in \{ \mathbf{U}, \mathbf{U}^\dagger \}$), and $\mathbf{h}_1$ (resp. $\mathbf{h}_2$) is a homodyne matrix interfering PNP outputs 1 with 2 and 3 with 4 (resp. 2 with 3 and 4 with 5). We use the PNP output 1 as an idler signal of known amplitude and phase to reconstruct the phase and amplitude of every other output in a cascaded manner (more details on the experimental setup is given in the Supplemental Information (SI), section I). 

Consequently, a single algorithm step of the INPRIS, shown in Figure \ref{fig:gen_concept}(b), is performed with four runs on the PNP in our proof-of-concept experiment (two unitary matrices, each requiring two homodyne measurements to reconstruct the amplitude and phase of the output vector). The required reconfiguration of the PNP phases for each of these runs, due to the state-dependent $\mathbf{R}[ S^{(\tau)}]$ and homodyne matrices $\mathbf{h}_{i}$, introduces an intrinsic noise on the measured outputs of the photonic system with standard deviation $\phi_\text{int}$, which depends on the single-shot fidelity of the encoded PNP transformations (the measured single-shot fidelity of the PNP was 91.6\% --- see SI, section II). We emphasize that, unlike our experimental demonstration of optical neural networks \cite{Shen2017a}, a deterministic PNP crosstalk correction step is performed just once before each PNP run, and that additional phase shifter optimization using a detection feedback loop is not required to optimize unitary fidelity. This high single-shot fidelity allows us to achieve low enough noise levels to probe Gibbs distribution in their ordered phase, while circumventing the increased complexity required to perform fidelity optimization at each iteration. 

This single-shot scheme mimics a passive or low-bandwidth photonic system, which motivates our treatment of non-idealities in the detected PNP outputs due to MZI phase settings as a separate intrinsic noise at every algorithm step. We also perturb the magnitude of each detected PNP output at the end of the algorithm step with an extrinsic zero-mean Gaussian noise with standard deviation $\phi_\text{ext}$ via the CPU (see Gaussian noise unit in Figure \ref{fig:setup}(c)). Assuming the intrinsic contribution to the noise is also Gaussian, these two independent sources of noise add to yield a total noise level $\phi = \sqrt{\phi_\text{int}^2 + \phi_\text{ext}^2}$. 

Figures \ref{fig:results}(b-f) show our main experimental results: the noise-dependent ground state population for various 4-spin Ising models. Each data point is averaged over 10 solver instances with randomized initial spin states and 100 algorithm steps per instance. By comparing simulation results (green curves correspond to ideal phase and amplitude reconstruction, while blue incorporate our experimental homodyne scheme) with larger number of iterations in Figures \ref{fig:results}(b-f), we observe that 100 algorithm steps is enough to converge to the ground state with high probability. In addition, we observe that our samples within the first 100 iterations show increased ground state probability over a random search (righthand plots in Figures \ref{fig:results}(b-f)). Since we can adjust the total noise in the system by tuning the extrinsic noise, $\phi_{\text{ext}}$, we can characterize the intrinsic noise level, $\phi_\text{int}$, by measuring the noise-dependent evolution of physical observables. Conventional physical observables describing spin glasses (energy, magnetization, heat capacity, susceptibility, etc.) can be calculated by using the thresholded PNP outputs as samples of the Gibbs distribution (Eq. (\ref{eq:gibbs})). Typically, these samples would be simulated using a heuristic algorithm \cite{Kirkpatrick1983, Landau2009, Aarts1988}, for instance, with conventional simulations run on a computer. The measured absolute magnetization of a 4-spin ferromagnet with dropout, measured by our PNP proof-of-concept, is shown in Figure \ref{fig:results}(a), which fits its theoretical value with a resulting intrinsic noise $\phi_\text{int} = 0.6457$.  The simulated values (green and blue plots) in Figures \ref{fig:results}(b-f) are calculated using the total noise level with the experimentally fitted $\phi_\text{int}$. This noise level naturally present in the photonic system can be leveraged to drive the ground state search as can be seen in Figures \ref{fig:results}(b-f), which show the ground state population for a variety of graphs (couplings shown in inset) as a function of the extrinsic noise standard deviation $\phi_\text{ext}$ applied in CPU. We notice that large ground state probabilities are attained for $\phi_\text{ext} = 0$, meaning that the regime of optimal noise level is attained with no further addition of extrinsic noise in the system. While noise is usually considered a nuisance in most physical (computing) systems, our experimental demonstration relies on physical noise, coming from finite single-shot fidelity of the PNP. This sheds a new light on noise as a computational resource in physical computing systems. 



When injecting extrinsic (CPU-applied) noise to the outputs at each algorithm step, we observe that the ground state probability remains large and agrees with simulations of the PNP that account for homodyne detection and phase-intensity reconstruction. The ideal performance of the PNP is shown in green for various numbers of algorithm steps per sampler instance, thus demonstrating the effect of our cascaded homodyne detection scheme  on the ground state convergence probability of the INPRIS. Overall, the ground state probability is $\geq 80\%$ for most graphs at some dropout level ($\alpha = 0$ or $\alpha = 1$) and low extrinsic noise levels, largely outperforming random search algorithms. We also observe that experiments \textit{with dropout} usually outperform those \textit{without dropout}, as expected from theory \cite{Roques-Carmes2018}. Extended experimental results on all generated graphs and dropout levels are available in the SI, section III.

\begin{figure}
\centering
\includegraphics[width = 0.35\textwidth]{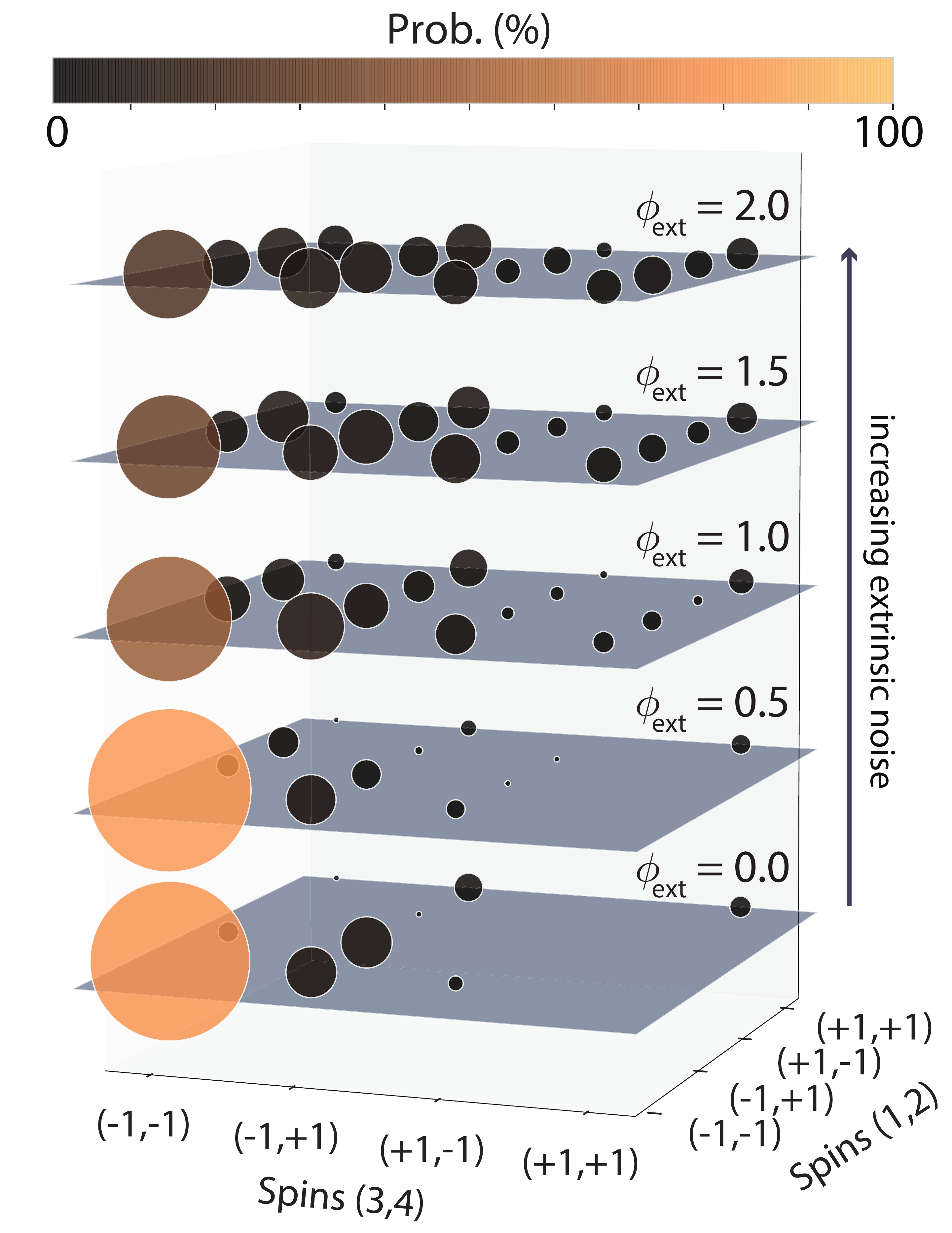}
\caption{\textbf{Phase space probing by extrinsic noise injection on ferromagnet problem.} The area of each dot is proportional to the measured mean probability of observing the system in state $(y,x)$ at a given extrinsic noise level $\phi_\text{ext}$, where $y = (\sigma_1, \sigma_2)$ is the state of spins 1 and 2, and $x = (\sigma_3, \sigma_4)$ is the state of spins 3 and 4.}
\label{fig:phasespace}
\end{figure}

We further investigated the impact of the homodyne reconstruction by probing the noise-dependent phase space of the two-dimensional Ising ferromagnetic problem, shown in Figure \ref{fig:phasespace}. The phase space population clearly shows a skew towards one of the ground states $\sigma = (-1,-1,-1,-1)$ (results are averaged over 10 random initial states). We confirm this skew originates from the cascaded homodyne detection by comparing the phase space population to simulations with (blue) and without (green) homodyne detections for the ferromagnetic graph (see SI, section IV). Since this skew breaks the two-fold degeneracy of the two-dimensional Ising problem, we presume that part of the phase information was lost through the reconstruction. To bypass this skew, we suggest scalable homodyne detection schemes which require more than one idler channels, in the SI, section I.  We observe that the skew does not seem to prevent the algorithm from converging to ground states but may complexify the evaluation of some physical observables. Interestingly, frozen transient states -- that can be eliminated by reducing the pump amplitude -- have also been observed in the OPO Ising machine \cite{Bohm2018}. As additional extrinsic noise is injected into the system, we observe that the phase space recovers its symmetry, thus providing an estimate for the amplitude of the skew in our photonic system. 


Fabrication imperfections in photonic networks could be a significant bottleneck in scaling integrated nanophotonic coherent Ising samplers to large ($N > 100$) graph orders. For instance, a static skew on the phase setting or the split ratio of beam splitters will result in a static error on the effective coupling between spins, thus reshaping the Hamiltonian landscape, which could impact the algorithm efficiency \cite{Roques-Carmes2018}. Thus, the reduction of imperfections is of paramount importance in the realization of the INPRIS on large-scale static photonic networks. However, several calibration techniques have been developed to achieve precise linear optical functions with broad process tolerances \cite{Miller2015a, Burgwal2017}. The bandwidth and energy cost of thermal phase shifters is another bottleneck in increasing the size of photonic networks. A significant advantage of the INPRIS over learning-based methods that require reconfiguration of the PNP phases at every algorithm step \cite{Roques-Carmes2018}, is that the phase reconfiguration bandwidth is irrelevant in characterizing the algorithm efficiency, since it only translates in a constant computational cost. Thus, a larger-scale INPRIS could be enabled by different architectures that do not suffer from this bottleneck and could operate at very low power levels. Examples include 3D-printed free-space optical networks \cite{Lin2018} and silicon photonic networks combined with non-volatile phase-change materials \cite{Wang2016}. 

To conclude, we introduce the INPRIS, a photonic circuit able to probe the Gibbs distribution of arbitrary Ising problems. We demonstrate a proof-of-concept experiment on a PNP that exhibits high success probability on a variety of 4-spin graphs. We show that noise coming from the PNP limited single-shot fidelity allows the system to operate close to its optimal noise regime, required to find ground states of Ising problems. Conversely, the influence of sources of noise of various origins (extrinsic vs. intrinsic) and natures (static skew vs. dynamic) is readily seen on physical observables such as magnetization, ground state population, and phase space distribution. In fully passive networks, other sources of noise should be leveraged. For instance, recent work on large scale photoelectric networks naturally demonstrate Gaussian noise on their outputs, arising at the quantum limit (and thus suggesting the operation of very large-scale INPRIS that reaches the thermodynamic limit, relevant for simulations in statistical mechanics with $N \sim 10^6$, that can reach attojoule energy consumption per algorithm unit step) \cite{Hamerly2018b}. Furthermore, the photonic circuit program for this architecture is fixed for any given coupling configuration. This feature enables the potential for quasi-passive photonic ASICs, such as non-volatile photonic ASICs \cite{Eggleton2011, deGalarreta2018, Wang2016}, that could deliver speed and energy savings over other physical Ising machines with active components.

This work also paves the way to larger-scale INPRIS by identifying key trade-offs in their design. While homodyne detection allows the reconstruction of the spin state phase and amplitude, an increased number of idler signals will increase the PNP footprint for a given number of spins and determine the overall accuracy of this operation. Other tradeoffs, such as optimal values of eigenvalue dropout \cite{Roques-Carmes2018}, should be taken into account when designing quasi-passive photonic GHz INPRIS on ASICs which could outperform current optical and electronic Ising machines by several orders of magnitude \cite{McMahon2016, Inagaki2016, Hamerly2018}. 

\section{Acknowledgments}
The authors would like to acknowledge Aram Harrow, Mehran Kardar, Ido Kaminer, Manan Raval, and Jamison Sloan for helpful discussions. This work was supported in part by the Semiconductor Research Corporation (SRC) under SRC contract \#2016-EP-2693-B (Energy Efficient Computing with Chip-Based Photonics -- MIT). This work was supported in part by the National Science Foundation (NSF) with NSF Award \#CCF-1640012 (E2DCA: Type I: Collaborative Research: Energy Efficient Computing with Chip-Based Photonics). This material is based upon work supported in part by the U. S. Army Research Laboratory and the U.S. Army Research Office through the Institute for Soldier Nanotechnologies, under contract number W911NF-18-2-0048. M. P. was financially supported by NSF Graduate Research Fellowship grant number 1122374. R. H. was supported by an IC Postdoctoral Research Fellowship at MIT, administered by ORISE through U.S. DOE and ODNI.  J.C. was supported by EU H2020 Marie Sklodowska-Curie grant number 751016.

\section{Author contributions}
C.R.-C., Y.S., D.E., and M.S. developed the original idea. M.P. and N.H. built the experimental setup. M.P., C.R.-C., Y.S., N.H., L.J., J.C., and R.H. performed the experiments. M.P. and C.R.-C. performed the data analysis. Y.S., J.D.J., D.E., and M.S. supervised the project. T.B.J. and M.H. fabricated the photonic integrated circuit. C.R.-C. and M.P. wrote the manuscript with inputs from all authors.

\section{Additional information}
Correspondence and requests for materials should be addressed to M.P., C. R.-C., and Y. S.

\section{Competing financial interests}
The authors declare the following patent application: U.S. Patent Application No.: 16/032,737. 

\vspace{-10pt}
\bibliographystyle{unsrt}
\bibliography{manuscript}

\end{document}


\title{A Recurrent Ising Machine in a Photonic Integrated Circuit \\ SUPPLEMENTAL INFORMATION}

\author{Mihika Prabhu$^{\bullet}$}  \email{mihika@mit.edu}
 \affiliation{MIT Research Lab of Electronics, 50 Vassar St, Cambridge, MA 02139, USA} 
 \author{Charles Roques-Carmes$^{\bullet}$} \email{chrc@mit.edu}
 \affiliation{MIT Research Lab of Electronics, 50 Vassar St, Cambridge, MA 02139, USA} 
\author{Yichen Shen$^{\bullet}$}\email{ycshen@mit.edu}
 \affiliation{Lightelligence, 268 Summer Street, Boston, MA 02210}
\author{Nicholas Harris}
 \affiliation{Lightmatter, 61 Chatham St 5th floor, Boston, MA 02109, USA}
\author{Li Jing} 
 \affiliation{MIT Department of Physics, 77 Massachusetts Avenue, Cambridge, MA 02139, USA} 
\author{Jacques Carolan} 
 \affiliation{MIT Research Lab of Electronics, 50 Vassar St, Cambridge, MA 02139, USA} 
\author{Ryan Hamerly} 
 \affiliation{MIT Research Lab of Electronics, 50 Vassar St, Cambridge, MA 02139, USA} 
\author{Tom Baehr-Jones} 
 \affiliation{Elenion Technologies, LLC, 171 Madison Ave 1100, New York, NY 10016, USA} 
\author{Michael Hochberg} 
 \affiliation{Elenion Technologies, LLC, 171 Madison Ave 1100, New York, NY 10016, USA} 
\author{Vladimir \v{C}eperi\'{c}} 
 \affiliation{MIT Department of Physics, 77 Massachusetts Avenue, Cambridge, MA 02139, USA} 
\author{John D. Joannopoulos} 
 \affiliation{MIT Department of Physics, 77 Massachusetts Avenue, Cambridge, MA 02139, USA} 
 \affiliation{Institute for Soldier Nanotechnologies, 500 Technology Square, Cambridge, MA 02139, USA \\ 
 $^{\bullet}$denotes equal contribution.}
 \author{Dirk R. Englund}
 \affiliation{MIT Research Lab of Electronics, 50 Vassar St, Cambridge, MA 02139, USA} 
\author{Marin Solja\v{c}i\'{c}}
 \affiliation{MIT Research Lab of Electronics, 50 Vassar St, Cambridge, MA 02139, USA} 
  \affiliation{MIT Department of Physics, 77 Massachusetts Avenue, Cambridge, MA 02139, USA}

\maketitle

\setcounter{tocdepth}{1}
\tableofcontents
\newpage

\section{PNP configuration setup}

In this section, we describe the programmable nanophotonic processor (PNP) main control steps. Here, we review the implementation on the PNP of the algorithm already presented in \cite{Roques-Carmes2018}. 
The PNP configuration can be decomposed in three main steps:
\begin{enumerate}
\item[$\triangleright$] \textit{General Initialization:} Generates an Ising model and initializes the different inputs to the PNP: algorithm parameters, initial spin state, homodyne detection matrices, etc.
\begin{enumerate}
\item[$\triangleright$] \textit{Inputs}
\begin{enumerate}
\item[$\bigstar$] \texttt{N} is the graph size, and \texttt{IsingType} is the type of graph (for example: \texttt{`Spin Glass'})
\item[$\bigstar$] \texttt{DD} is the eigenvalue dropout level 
\item[$\bigstar$] \texttt{AlgIdler} is the value of the idler signal in algorithm space.
\end{enumerate}
\item[$\triangleright$] \textit{Outputs}
\begin{enumerate}
\item[$\bigstar$] \texttt{T, Hmin} are the Ising matrix and the ground state energy: $H_{\text{min}} = \min_{\sigma} -\frac{1}{2}  \sum_{i,j}\sigma_i T_{i,j} \sigma_j$.
\item[$\bigstar$] \texttt{K, U, D0, J, D1, C} are matrices used in the decomposition of the Ising problem. $K = U D_0 U^\dagger$ where $K$ has been added a diagonal offset of amplitude depending on \texttt{DD}. $J = U \sqrt{D_0} U^\dagger$ and $C = 2J$.
\item[$\bigstar$] \texttt{Thresholds} is the threshold vector used in the algorithm (to define the nonlinear operation).
\item[$\bigstar$] \texttt{h1, h2} are homodyne detection matrices.
\item[$\bigstar$] \texttt{S, spins, In, H} are initial reduced spins, spins, inputs states and energy.
\end{enumerate}
\end{enumerate}

\item[$\triangleright$] \textit{Configure PNP:} Feeds the PNP with phases parameters, given an input and Ising problem. This function outputs a \texttt{.txt} file that can be read by the nanophotonic processor. 
\begin{enumerate}
\item[$\triangleright$] \textit{Inputs}
\begin{enumerate}
\item[$\bigstar$] \texttt{Input} is the input in algorithm space (spins). It is the concatenation of a spin state of size $N$ with the idler value in spin state (\texttt{AlgIdler==1}).
\item[$\bigstar$] \texttt{InputField} is the electric field input fed to the PNP, of the form $(E_0,0,0,0,0)^T$.
\item[$\bigstar$] \texttt{U} is the unitary matrix used in the diagonalization of the Ising problem.
\item[$\bigstar$] \texttt{h1, h2} are the homodyne detection matrices.
\item[$\bigstar$] \texttt{ChipIndex, kChip} are respectively the index of the current PNP $\in (1,2)$ and the number round trips so far.
\end{enumerate}
\end{enumerate}

\item[$\triangleright$] \textit{Output reader:} Given data from two homodyne detection, reconstructs output of the matrix multiplication in algorithm space. Also applies electronic feedback operations: diagonal multiplication, digital noise, nonlinear threshold, computes energy, etc.
\begin{enumerate}
\item[$\triangleright$] \textit{Inputs}
\begin{enumerate}
\item[$\bigstar$] \texttt{Out1, Out2} are the two detected output from the detector
\item[$\bigstar$] \texttt{D1, Thresholds, T, AlgIdler} are defined above.
\item[$\bigstar$] \texttt{DIGITAL\_NOISE\_FLAG} is a boolean defined whether extrinsic noise should be added to the output. \texttt{Phi} is the digital noise level.
\item[$\bigstar$] \texttt{ChipIndex} is the chip index $\in (1,2) $
\end{enumerate}
\item[$\triangleright$] \textit{Outputs}
\begin{enumerate}
\item[$\bigstar$] \texttt{NextFeed, NextS, Nextspins} are the next input, reduced spin and spin states.
\item[$\bigstar$] \texttt{NextH} is the new energy.
\end{enumerate}
\end{enumerate}

\end{enumerate}

\subsection{Initialization and first feed-in}
We first generate all relevant variables for a given problem with the \textit{General Initialization}. The outputs of this function need to be global parameters as they will be useful throughout the run of the algorithm. Then, we use the current input \texttt{In(:,1)} and \texttt{ChipIndex==1} to generate phase parameters to code for the unitary matrices $h_1 U^{\dagger} R_{\mathtt{In(:,1)}}$ and $h_2 U^{\dagger} R_{\mathtt{In(:,1)}}$ where $h_1, h_2$ are two homodyne detection matrices ($h_1$ (resp. $h_2$) interferes outputs $i$ and $i+1$ starting with $i=1$ (resp. $i=2$)), $R_{\mathtt{In(:,1)}}$ is a unitary matrix rotating $(1,0,...,0)^T$ into \texttt{In(:,1)} and $U^\dagger$ is the unitary matrix present in the eigenvalue decomposition of our Ising weight matrix. Phases are generated assuming we can tune an $N\times N$ array of MZIs arranged in a rectangular pattern.
This algorithm needs to be run twice on the PNP (corresponding to two homodyne detections):
\begin{enumerate}
\item[1.] Configure the PNP with phases \texttt{MZI\_Array11}. We save the detected output as \texttt{Out11}.
\item[2.] Configure the PNP with phases \texttt{MZI\_Array12}. We save the detected output as \texttt{Out12}.
\end{enumerate}

\subsection{Output reading of PNP round 1, feed-in PNP round 2}
We detect \texttt{Out11} and \texttt{Out12}, then to convert detected data into \texttt{In(:,2)}. Then, we configure the PNP with this new input \texttt{In(:,2)} and \texttt{ChipIndex==2}.
\begin{enumerate}
\item[1.] Calibrate PNP with phases \texttt{MZI\_Array21}. We save the detected output as \texttt{Out21}.
\item[2.] Calibrate PNP with phases \texttt{MZI\_Array22}. We save the detected output as \texttt{Out22}.
\end{enumerate}
 
\subsection{Output reading of PNP round 2, feed-in PNP round 3}
We detect \texttt{Out21} and \texttt{Out22}, then convert detected data into \texttt{In(:,3)} (data in algorithm space). Then, we configure the PNP with this new input \texttt{In(:,3)} and \texttt{ChipIndex==1}.

\subsection{Recurrent step}
Every algorithm step translates into 4 runs on the PNP, corresponding to two unitary matrices, and two homodyne detections per unitary matrix.
\begin{enumerate}
\item[$\triangleright$] Round-trip on PNP 1, coding for $U^\dagger$
\begin{enumerate}
\item[$\triangleright$] Homodyne detection 1 : $h_1 U^\dagger R_{\mathtt{In(:,2*kChip+1)}}$
\item[$\triangleright$] Homodyne detection 2 : $h_2 U^\dagger R_{\mathtt{In(:,2*kChip+1)}}$
\end{enumerate}
\item[$\triangleright$] Round-trip on PNP 2, coding for $U$
\begin{enumerate}
\item[$\triangleright$] Homodyne detection 1 : $h_1 U R_{\mathtt{In(:,2*kChip+2)}}$
\item[$\triangleright$] Homodyne detection 2 : $h_2 U R_{\mathtt{In(:,2*kChip+2)}}$
\end{enumerate}
\end{enumerate}

\subsection{Homodyne detection algorithm}
In the main text, we compare the experimental results to two kinds of simulation. The simulation refered to as "ideal phase-intensity reconstruction" is the recurrent transformation dicussed in \cite{Roques-Carmes2018}. For the "homodyne phase-intensity reconstruction", we also simulate the propagation of light on-chip through the MZI array. The two homodyne detection matrices we use are:
\begin{eqnarray}
h_1 = \begin{bmatrix} 
    s & s & 0 & 0& 0 \\
    -s & s&  0& 0&0 \\
    0 & 0 & s & s&0 \\
	0 & 	0 & -s & s&0   \\
	0 & 0 & 0& 0& 1  
    \end{bmatrix} \ \ \ \ 
    h_2 = \begin{bmatrix} 
    1 & 0 & 0 & 0& 0 \\
    0 & s&  s& 0&0 \\
    0 & -s & s & 0&0 \\
	0 & 	0 & 0 & s&s   \\
	0 & 0 & 0& -s& s  
    \end{bmatrix}
\end{eqnarray}
where $s = 1/\sqrt{2}$. In order to reconstruct the real output $x$ from the two homodyne detection outputs, we apply the following cascaded algorithm. 

\begin{algorithm}
\begin{algorithmic}
\STATE Input data $x$ to reconstruct 
\STATE $y_1 \leftarrow h_1 x$
\STATE $y_2 \leftarrow h_2 x$
\STATE $\text{Ref} \leftarrow 1$
\STATE $x_r(i) \leftarrow  y_2(1)$ (reconstructed $x$ vector) 
\FORALL{$i \in \{2, ..., N\}$}
\IF{$i$ is even}
\STATE $d \leftarrow y_1(i-1) - y_1 (i)$
\ENDIF
\IF{$i$ is odd}
\STATE $d \leftarrow y_2(i-1) - y_2 (i)$
\ENDIF
\\
$x_r(i) \leftarrow \frac{d}{2 \text{Ref} x_r(i-1)}$
\ENDFOR
\end{algorithmic}
\caption{Cascaded homodyne algorithm for PNP output reconstruction.}
\label{homo_algo}
\end{algorithm}

A potential issue with the above algorithm is the cascading of error throughout the reconstruction of the output. Also, if $x_r(i) =0$, then in theory $x_r(i+1)$ diverges. We suggest a more general class of algorithms in order to prevent these divergences, assuming we have access to more channels. Since we want to extract $2N$ variables (the phase \textit{and} amplitude of $N$ PNP outputs), the reconstruction algorithm requires $2N$ measurements. Let us assume we have access to $1\leq k \leq N$ local oscillators of known amplitude and phase, in order to interfere them with the $N$ PNP outputs.
\begin{enumerate}
\item[$\triangleright$] If $k=1$, the algorithm is similar to Algorithm \ref{homo_algo}. 
\item[$\triangleright$] If $k=N$, the issue of algorithmic divergence is avoided. However, the effective footprint of the PNP is doubled compared to the one required to implement Algorithm \ref{homo_algo}.
\item[$\triangleright$] Thus, by choosing $1 <k < N$, we can find a middle ground between optimizing the chip footprint and the issue of algorithmic divergences. 
\end{enumerate}


\subsection{Decomposition Algorithm}
To perform an arbitrary unitary matrix multiplication on-chip, we need to decompose the matrix into a product of $2\times 2$ unitary transformations, arranged in a lattice that corresponds to the shape of our MZI array. The unit block of our decomposition corresponds to a transformation through a single Mach-Zehnder Interferometer:
\begin{equation}
   T_{n,n+1} (\theta, \phi) = \left[ {\begin{array}{cc}
   e^{i\phi} & 0 \\
   0 & 1 \\
  \end{array} } \right] \frac{\sqrt{2}}{2} \left[ {\begin{array}{cc}
   1 & i \\
   i & 1 \\
  \end{array} } \right] \left[ {\begin{array}{cc}
   e^{i\theta} & 0 \\
   0 & 1 \\
  \end{array} } \right] \frac{\sqrt{2}}{2} \left[ {\begin{array}{cc}
   1 & i \\
   i & 1 \\
  \end{array} } \right] = i e^{i\theta/2}
  \left[ {\begin{array}{cc}
   e^{i\phi} \sin \left( \frac{\theta}{2} \right) & e^{i\phi}  \cos \left( \frac{\theta}{2} \right) \\
   \cos \left( \frac{\theta}{2} \right) & -\sin \left( \frac{\theta}{2} \right) \\
  \end{array} } \right]
  \label{MZI_eq}
\end{equation}
which applies on channels $(n,n+1)$. The rest of the channels go through an identity transformation. We modify the original decomposition algorithm on a square lattice \cite{Clements2016} for our application as follows:

\begin{algorithm}[H]
\begin{algorithmic}
  \FORALL{$i$ from $1$ to $N-1$}
  	\IF{$i$ is odd}
	  	   \FORALL{$j$ from $0$ to $i-1$}
 			\STATE $U_{N-j,i-j} := a_r + i a_i$ 
	 		\STATE $U_{N-j,i-j+1} := b_r + i b_i$ 
 			\STATE $\phi \leftarrow \pi-\mathrm{atan} \frac{a_i b_r -a_r b_i}{a_r b_r+a_i b_i}$
			\STATE $\theta \leftarrow  2 \mathrm{acot} \frac{-a_r \cos \phi + a_i \sin \phi}{b_r} $
			\STATE $V \leftarrow V T_{i-j,i-j+1}(\theta,\phi)$
		\ENDFOR
	\ENDIF
	  	\IF{$i$ is even}
		   \FORALL{j from $1$ to $i$}
			\STATE $U_{N+j-ii-1,j} := a_r + i a_i$ 
			\STATE $U_{N+j-i-1,j} := b_r + i b_i$ 
			\STATE $\phi \leftarrow \mathrm{atan} \frac{a_i b_r -a_r b_i}{a_r b_r+a_i b_i}$
			\STATE $\theta \leftarrow  2 \mathrm{atan} \frac{a_r \cos \phi+ a_i \sin \phi}{b_r} $
			\STATE $V \leftarrow T^{-1}_{N+j-i-1,N+j-i}(\theta,\phi) V$
			\ENDFOR
	\ENDIF
\ENDFOR
 \end{algorithmic}
\caption{Algorithm to decompose a unitary transformation $V$ into our MZI lattice.}
\end{algorithm}
The output of this algorithm is a decomposition that has the following form: 
\begin{equation}
U = \Big( \prod_{(m,n) \in S_L} T_{m,n} \Big) D \Big( \prod_{(m,n) \in S_R} T_{m,n}^{-1} \Big) 
\end{equation}
where $D$ is a diagonal of phase shifts. To finish the decomposition, we use the fact that for every unitary transformation $T$, there exists $D'$ such that $D T^{-1}(\theta, \phi) = T (\theta', \phi')D'$, with the following conditions on the different phases:
\begin{eqnarray}
\theta' &=& \theta \\
\phi' &=& \phi_1 - \phi_2 \\
\phi_1' &=& \phi_2 - \phi - \pi - \theta \\
\phi_2' &=& \phi_2 - \pi - \theta 
\end{eqnarray}
where $D = \left[ {\begin{array}{cc}
   e^{i\phi_1} & 0 \\
   0 & e^{i\phi_2} \\
  \end{array} } \right]$ and $D' = \left[ {\begin{array}{cc}
   e^{i\phi_1'} & 0 \\
   0 & e^{i\phi_2'} \\
  \end{array} } \right]$.
  We thus get the following decomposition in the end: 
\begin{equation}
U = \Big( \prod_{(m,n) \in S_L} T_{m,n} \Big) \Big( \prod_{(m,n) \in S_R} T_{m,n}^{-1} \Big) D'
\end{equation}
  
\section{PNP configuration and calibration}
\subsection{Experimental setup configuration}
\begin{figure}
\centering
\includegraphics[width=0.7\textwidth]{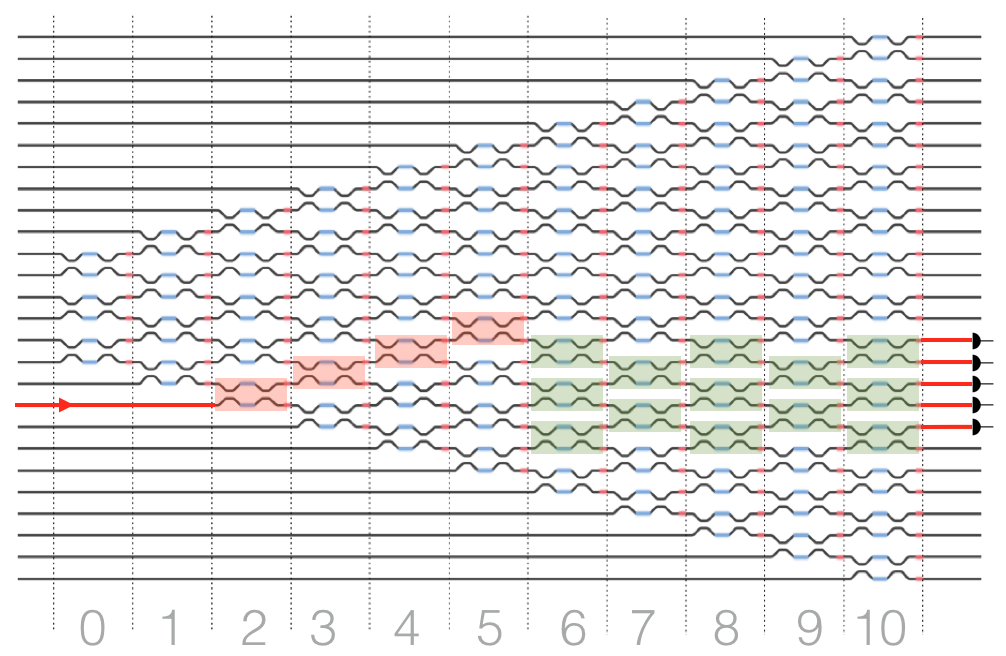} 
\caption{Laser-written glass interposer fanout structure manufactured by TEEM Photonics for coupling between 127$\mu$m pitch telecom fiber array and 25.4$\mu$m pitch chip facets.}
\label{fig:expt_setup}
\end{figure}

The programmable nanophotonic processor (PNP) chip used in our experiments is a 26-mode silicon-on-insulator photonic integrated circuit fabricated by the OpSIS foundry. Coherent continuous-wave light at 1550nm, rotated into TE polarization, is coupled to the input of the PNP chip via a laser-written glass interposer fanout chip. These interposer chips, fabricated by TEEM Photonics, interface standard polarization-maintaining SMF telecommunication fiber (10$\mu$m mode field diameter)  to the 2$\mu$m mode field diameter of the on-chip edge couplers and have characterized insertion losses in the range [0.47dB, 0.87dB] per channel. The optical signal vector is encoded in the amplitude and phase of guided light within the 26 on-chip waveguide modes. To accommodate space constraints and avoid global phase differences on the input, a single input port is excited and routed (shown in red, Fig. \ref{fig:expt_setup}) to a 5x5 program unit (shown in green, Fig. \ref{fig:expt_setup}). The program unit encodes a unitary matrix product of the state preparation, Ising unitary, and homodyne detection matrices, as detailed in Section I.1. 

The resultant program unit outputs are routed to an array of InGaAs p-i-n photodiodes with responsivities of approximately 0.93 A/W for 1550nm and a 2GHz cutoff frequency. Intensity readout is limited, in practice, to approximately 4kHz by the rise time of the readout circuitry.  The noise floor of the detectors is observed to be on the order of 2nW, with a saturation power of 20$\mu$W.  In order to counteract errors resulting from the division step in the homodyne reconstruction algorithm (Section I.5), PNP outputs are only accepted if the total output power of the signal modes exceeds 1$\mu$W.

Coupling efficiency for the Gaussian waveguide beams to and from the chip decreases exponentially with the degree of misalignment between the fiber array and the chip facet. Once coupled, this misalignment can often occur due to the difference in the thermal expansion coefficients between the silicon chip and the glass interposer. To maintain stable coupling in the presence of thermal fluctuations from the environment and the heat-dissipating phase shifters on the chip, it is necessary to implement a thermal control system to stabilize the temperature on the PNP.  We mount the chip and a thermistor on a copper block with thermally conductive paste. The thermistor measures the temperature of the chip and, in combination with a Peltier cooler glued to the copper block, a p-i-d feedback loop is established with an Arroyo Instruments TEC temperature controller to maintain the on-chip temperature to within 0.01 Kelvin.

\subsection{PNP calibration}
Nonidealities including imperfections in fabricated waveguides and inaccuracy in voltage control can severely degrade single-shot fidelity of the optical matrix multiplication unit (OMMU).  Motivated by decomposition schemes mapping an arbitrary unitary operator to an array of Mach-Zehnder interferometer (MZI) phases \cite{Clements2016, Reck1994}, we extract the phase-voltage relationship for each voltage-controlled thermal phase shifter on the chip using an MZI calibration protocol detailed in \cite{Harris2016}.



\begin{figure}[h]
\centering
\includegraphics[width=0.3\columnwidth]{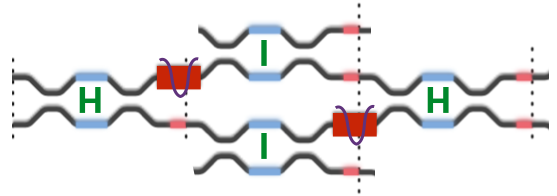} 
\caption{Meta-MZI configuration. The combination of four single MZIs forms a meta-MZI when the top/bottom MZIs are programmed to be identities and the left/right MZIs are programmed to be Hadamard transforms. The internal phase difference of the meta-MZI can be tuned by varying the phase between the external phase shifters of the left and bottom MZIs. }
\label{fig:meta_mzi}
\end{figure}

It is straightforward to measure an intensity fringe that depends on an MZI's internal phase shifter, $\theta$; however, we observe from the MZI transfer matrix shown in Equation (\ref{MZI_eq}) with internal phase difference, $\theta$, and external phase  difference, $\phi$, that the intensities detected at the two output modes of an MZI are independent of $\phi$. Nonetheless, accurate setting of the external phase shifters is crucial to the implementation of unitary matrices on the PNP when there are multiple MZI layers in the decomposition ~\cite{Reck1994, Clements2016}. One can still observe an intensity signature from the external phase shifters by programming four adjacent MZIs in the configuration shown in Figure \ref{fig:meta_mzi}.  The left and right MZIs are programmed as symmetric beam  splitters that implement the Hadamard operation \cite{NielsenChuang},
\begin{eqnarray}
H = \frac{1}{\sqrt{2}}
\begin{bmatrix}
1 & 1 \\
1 & -1 \\
\end{bmatrix}
\end{eqnarray}
while the top and bottom MZIs are identities, thereby generating a meta-MZI that behaves like a traditional MZI where the internal phase difference of the meta-MZI can be tuned by varying the external phase shifters of the left and bottom MZIs.

\begin{figure}[h]
\centering
\includegraphics[width=0.9\columnwidth]{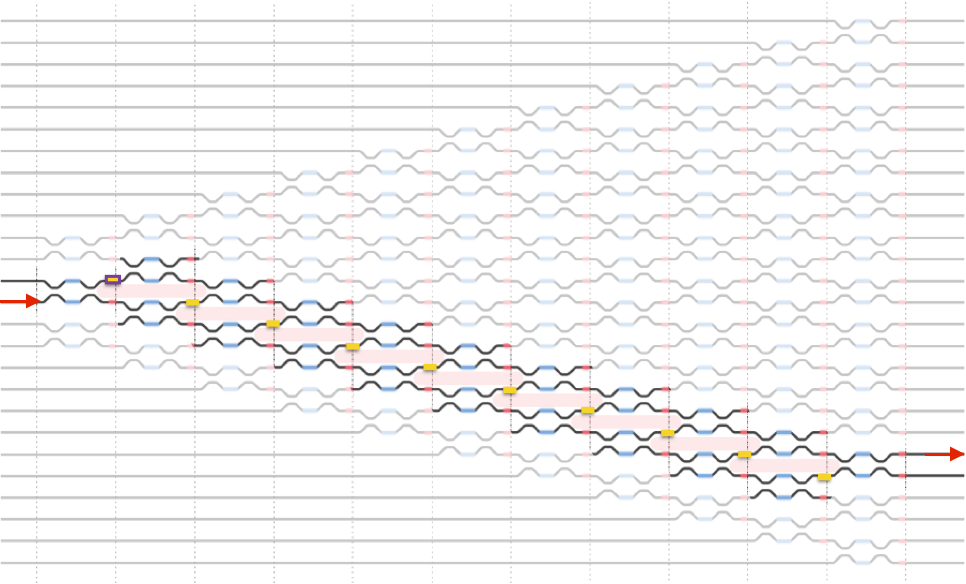} 
\caption{Meta-MZI calibration ``chain". External phase shifter voltages are swept to generate transmission fringes for the entire chain of meta-MZIs, which can be processed to determine external phase vs. voltage relationship and offsets.}
\label{fig:meta_cal}
\end{figure}

Given an individual meta-MZI, we seek to zero out the relative phase difference between the meta-MZI's internal arms as a result of the left and bottom individual MZI external phase shifters. In order to do this, we sweep the individual MZI external phase shifters (shown as  yellow blocks in Figure \ref{fig:meta_cal}) until the given meta-MZI performs a ``swap'' operation. The ``swap" configuration indicates that there is zero phase difference between the internal arms of the meta-MZI.

For each meta-MZI, there are two tunable phase shifters and two constant phase shifts inherent to the fabricated structure that, together, produce the net phase difference between the meta-MZI's internal arms. In addition, adjacent meta-MZIs share one phase shifter. Relating the swap phase condition for neighboring meta-MZIs results in a system of $m$ coupled equations in $m$ variables, where each chain of meta-MZIs has one phase shifter (denoted in Figure \ref{fig:meta_cal} by the top yellow box outlined in purple) that serves as a reference phase, set to $\pi$. We calculate the number of non-reference external phase shifters that we must calibrate to be given by
\begin{eqnarray}
m = (\text{\# meta-MZIs}) - (\text{\# meta-MZI "chains"})  = 63-11 = 52
\end{eqnarray}

Furthermore, for each meta-MZI, the swap condition necessitates the following relationship between the tunable and intrinsic phases:
\begin{eqnarray}
\phi_{\text{top}} - \phi_{\text{bottom}} + \phi_{\text{config,diff}} + \Delta = 0
\end{eqnarray}
where $\phi_{\text{top}}$ and $\phi_{\text{bottom}}$ are the external phase shifters for the left and bottom individual MZIs, respectively, $\phi_{\text{config,diff}}=\pi$ (implementation of the identity requires a phase difference of $(\theta, \phi) = (\pi,\pi)$ for the top and bottom individual MZIs), and $\Delta$ is the inherent phase difference between the two arms from variations in the waveguide fabrication.

The inherent phase difference, $\Delta$, is fitted from the transmission sweeps of both $\phi_\text{top}$ and $\phi_\text{bottom}$ for a given meta-MZI. Using the  fitted $\Delta$ intrinsic phase difference for each meta-MZI, we solve a simple matrix equation to determine the phase shifter offsets that cancel out the effect of fabrication imperfections on the external phase shifters.  We also use these fitted transmission sweeps to extract a phase vs. voltage relationship for each individual external phase shifter. 

\subsection{PNP characterization}
Given the calibrated phase vs. voltage functions for each of the individual phase shifters on the PNP, we seek to determine how accurately we can implement a desired unitary transform, with elements shown below:
\begin{eqnarray}
U = 
\begin{bmatrix}
r_{11}e^{i\theta_{11}} & r_{12}e^{i\theta_{12}} & \cdots & r_{1n}e^{i\theta_{1n}} \\
r_{21}e^{i\theta_{21}} & r_{22}e^{i\theta_{22}} & \cdots & r_{1n}e^{i\theta_{2n}} \\
\vdots & \vdots & \ddots & \vdots \\
r_{n1}e^{i\theta_{n1}} & r_{n2}e^{i\theta_{n2}} & \cdots & r_{nn}e^{i\theta_{nn}} \\
\end{bmatrix}
\end{eqnarray}

\begin{figure}[h]
\centering
\includegraphics[width=0.6\columnwidth]{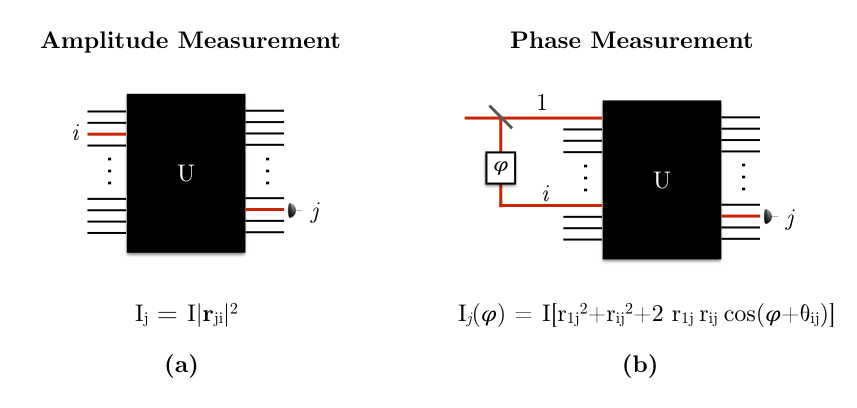} 
\caption{Unitary matrix characterization protocol. (a) Magnitudes are measured using an intensity measurement at the detectors. (b) Phases are deduced from fitting the interference fringe at a selected output that results from two input modes being simultaneously illuminated, with a variable phase difference between the input arms.}
\label{fig:unitary_char}
\end{figure}

We measure the magnitudes, $r_{ij}$ and phases, $\theta_{ij}$, using two different protocols (outlined in Figure ~\ref{fig:unitary_char}). 

We measure the amplitude $r_{ij}$ by measuring the output power on the $j^{th}$ detector $I_{j}$ upon illumination of input $i$ by with intensity $I$:
\begin{eqnarray}
|r_{ij}| = \sqrt{\frac{I_{j}}{I}}
\end{eqnarray}

Measurement of $\theta_{ij}$ is slightly more involved, as interference is required to observe an intensity signature from these phase terms.  This interference is implemented using a protocol developed by \cite{RahimiKeshari2013} (outlined in Fig \ref{fig:unitary_char}). We insert a coherent state input to the chip, program the MZI in the first layer to be a 50:50 beamsplitter,  route the two split coherent states to input mode 1 and $i$ of the unitary matrix circuit, and measure from output mode $j$ while sweeping the value of a tunable phase difference, $\phi$, between the split coherent states.  Because the matrix is unitary and the columns and rows form a unitary basis, we can assume, without loss of generality, that the first column and first row of the characterized matrix are real \cite{RahimiKeshari2013}.  We then use these ``border'' elements as references upon which to calculate the rest of the phases $\theta_{ij}$.  We first find the value of $\phi$ that maximizes the intensity fringe generated at output $j$  when we sweep $\phi$. Fitting the resultant fringe with the following relation subsequently gives us the value of $\theta_{ij}$
\begin{eqnarray}
I_{j}(\phi) = I\left|r_{1j} + r_{ij}e^{i(\phi + \theta_{ij})}\right|^2 = I[r_{1j}^2 + r_{ij}^2 + 2r_{1j}r_{ij}\cos(\phi + \theta_{ij})]
\end{eqnarray}

\subsection{PNP voltage crosstalk correction}
\begin{figure}[h]
\centering
\includegraphics[width=0.6\columnwidth]{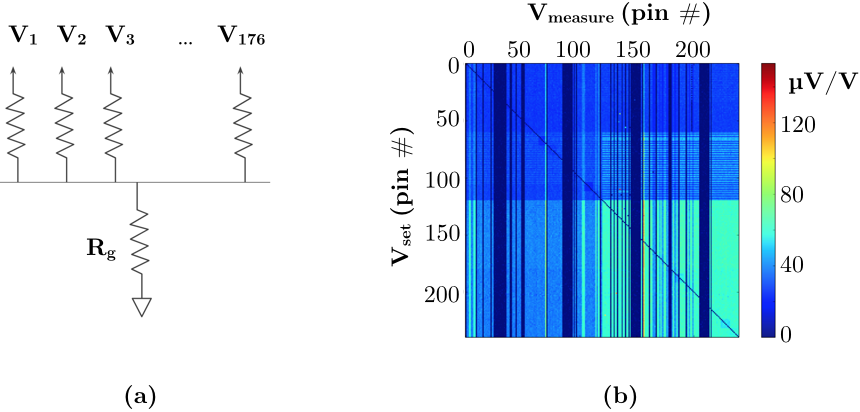} 
\caption{(a) Voltage crosstalk model. A small ground resistance produces voltage crosstalk that couples the voltages applied by all of the pins. (b) Measured linear correlation matrix quantifying crosstalk between all 240 pins.}
\label{fig:v_xtalk}
\end{figure}

Programming an N-dimensional unitary matrix requires at least $(N)(N-1)$ phase shifters to be programmed simultaneously.  When programming many MZIs at once, it is necessary to consider non-idealities in the structure of the resistive network driving the thermo-optic phase shifters in order to correct for voltage crosstalk in the electronic circuitry (shown in Figure \ref{fig:v_xtalk}a). In particular, the ground terminal comprises of a network of physical metal pours both on the PNP chip and on the control PCB, leading to a small resistance between the heater grounds and true ground. With each heater that is turned on, a small current flows through the ground resistor, changing the effective voltage drop across all the other resistors in the network.

To characterize the effect of this ground resistance on the 240 pins of the driver+ board, we measure the 240-dimensional crosstalk coupling matrix (shown in Figure \ref{fig:v_xtalk}b). For each of the 240 voltage driver pins, a single pin was probed while each of the other 239 pins was swept in voltage between 0 Volts and 7.0 Volts in 10 increments.  The resulting voltage crosstalk curves are not perfectly linear, however, so we apply a polynomial correction model described in the following paragraph, and outlined further in \cite{Prabhu2018}.

We observe increasingly nonlinear behavior in crosstalk curves at higher voltages, possibly due to the change in heater resistance due to decreased carrier mobility at higher temperatures \cite{Lundstrom2009}. We, therefore, model each measured voltage crosstalk curve as a cubic function of the set voltages, $\vec{V}_{\text{set}}$: 
\begin{eqnarray}
\vec{V}_{\text{meas}} = \overline{\overline{\mathbf{C}}}^{(1)}\vec{V}_{\text{set}} +  \overline{\overline{\mathbf{C}}}^{(2)}\vec{V}_{\text{set}}^2 +  \overline{\overline{\mathbf{C}}}^{(3)}\vec{V}_{\text{set}}^3
\end{eqnarray}
and use Newton's method \cite{Kelley1995} to find a solution vector, $\vec{V}_{\text{set}}$ for this nonlinear system of equations that produces a desired measured voltage vector.  

To test the effect of this protocol, a sample of 100 unitary matrices were generated at random from using the following procedure: 
\begin{eqnarray}
\mathbf{C} &&= \text{random complex NxN matrix} \\
\mathbf{H} &&= \frac{1}{2} (\mathbf{C} + \mathbf{C}^{\dagger}) \\
\mathbf{U} &&= e^{(i\mathbf{H})}  
\end{eqnarray}

\begin{figure}[h]
\centering
\includegraphics[width=0.4\columnwidth]{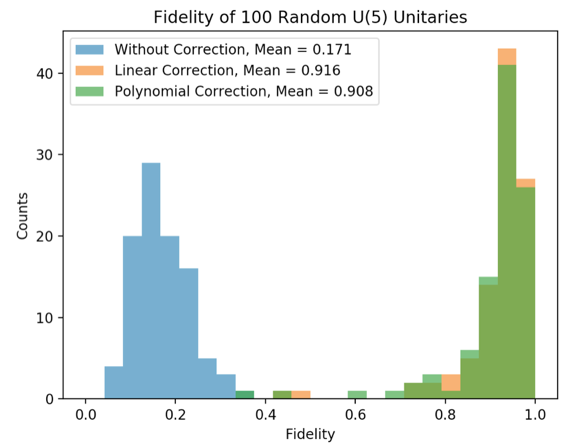} 
\caption{Fidelities of 100 randomly generated U(5) unitary matrices with and without voltage crosstalk correction. Polynomial crosstalk correction assumes a cubic voltage crosstalk relation, solved using Newton's method. Linear crosstalk correction assumes a linear relation, enabling a solution to be obtained by inverting the crosstalk correlation matrix. }
\label{fig:fidelities}
\end{figure}

After use of the characterization scheme outlined in Section II.3, we calculate the fidelity of the programmed matrices using the following unitary fidelity metric, with results shown in Figure \ref{fig:fidelities}.
\begin{eqnarray}
F = \frac{|\text{Tr}(\mathbf{U}_{m}^{\dagger}\mathbf{U})|}{N}.
\end{eqnarray}

\section{Extended results on ground state search}

\begin{figure}
\begin{center}
\includegraphics[scale=0.4]{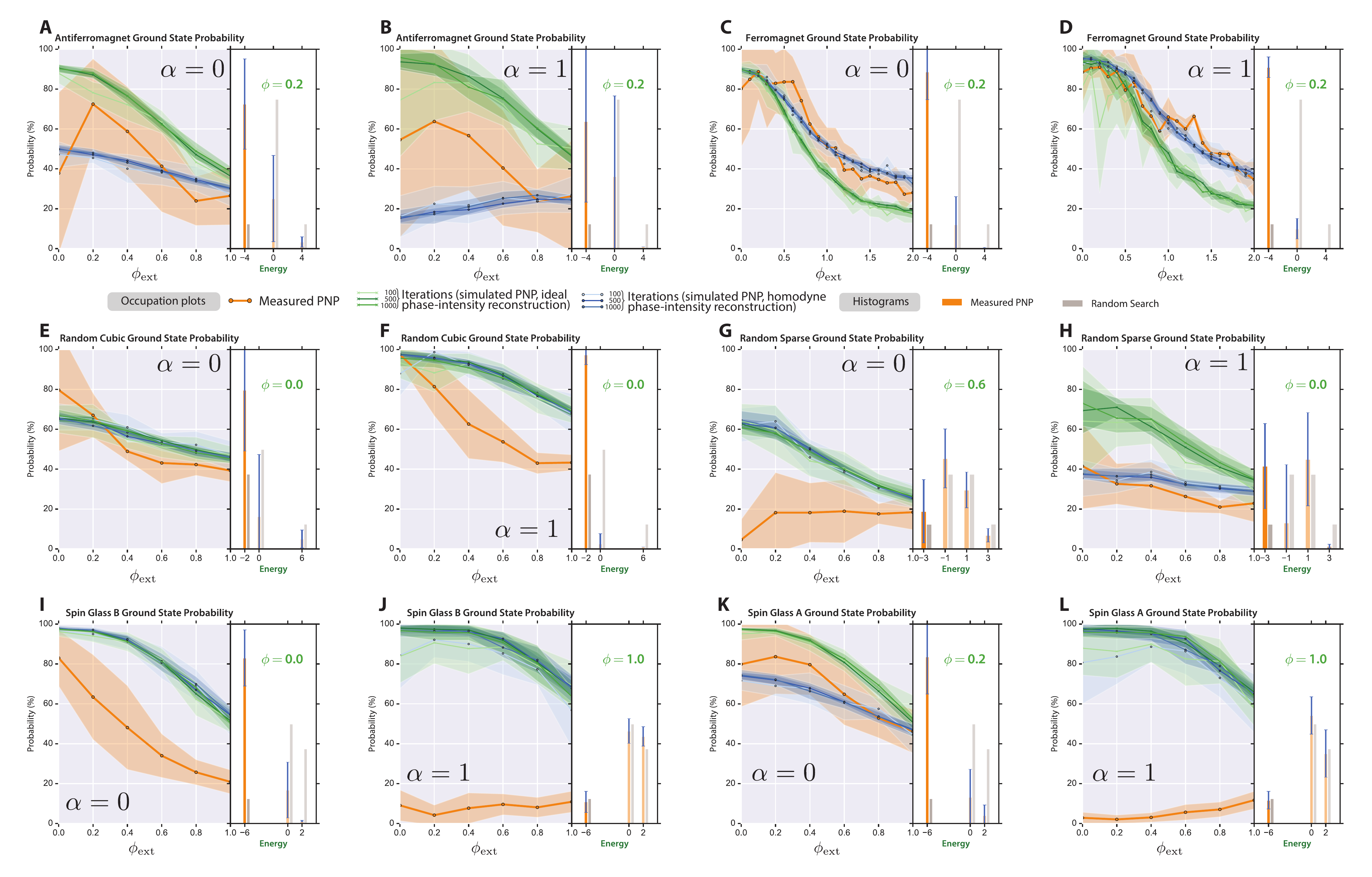}
\caption{\textbf{Extended results on ground state search.} Population plots and histograms are shown for antiferromagnet \textbf{(A, B)}, ferromagnet, \textbf{(C, D)}, random cubic graph \textbf{(E, F)}, random sparse graph \textbf{(G, H)}, spin glass B \textbf{(I, J)}, and spin glass A \textbf{(K, L)}. \textbf{(A, C, E, G, I)} correspond to dropout ($\alpha = 0$) and \textbf{(B, D, F, H, J, L)} to no dropout ($\alpha = 1$).}
\label{all_probs}
\end{center}
\end{figure}

The dependence of the ground state population as a function of the extrinsic noise is plotted for all studied graphs and two levels of dropout $(\alpha = 0$ and $\alpha = 1$) in Figure \ref{all_probs}.

\section{Phase space exploration characterization}

In this section, we represent the Ising phase space for a variety of graphs and extrinsic noise level $\phi = 1.0$. In each plot, the $y$ (resp. $x$) coordinate corresponds to the binary representation of the spins $(\sigma_1, \sigma_2) \in \{ (-1,-1), (-1,+1), (+1,-1), (+1,+1) \}$ (resp. $(\sigma_3, \sigma_4)$). The area of each dot is proportional to the probability of spin state $(y, x)$, which is also encoded in the color of the dot. At each coordinate, three concentric dots are plotted, with respective areas proportional to the mean probability, the mean probability minus its standard deviation, and the mean probability plus its standard deviation (all statistics for each plot are averaged over 10 runs of the same graph with random initial states).

\begin{figure}
    \centering
    \begin{subfigure}[t]{\textwidth}
        \includegraphics[width=\textwidth]{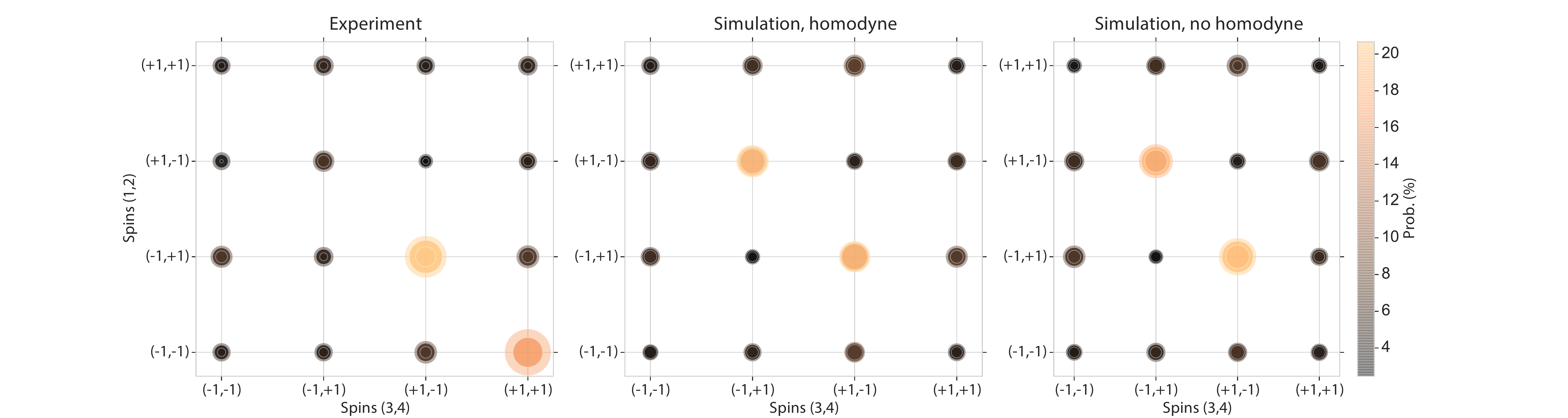}
        \caption{$\alpha = 0$, antiferromagnet}
        \label{fig:antiferro_DD0}
    \end{subfigure}
    \begin{subfigure}[t]{\textwidth}
        \includegraphics[width=\textwidth]{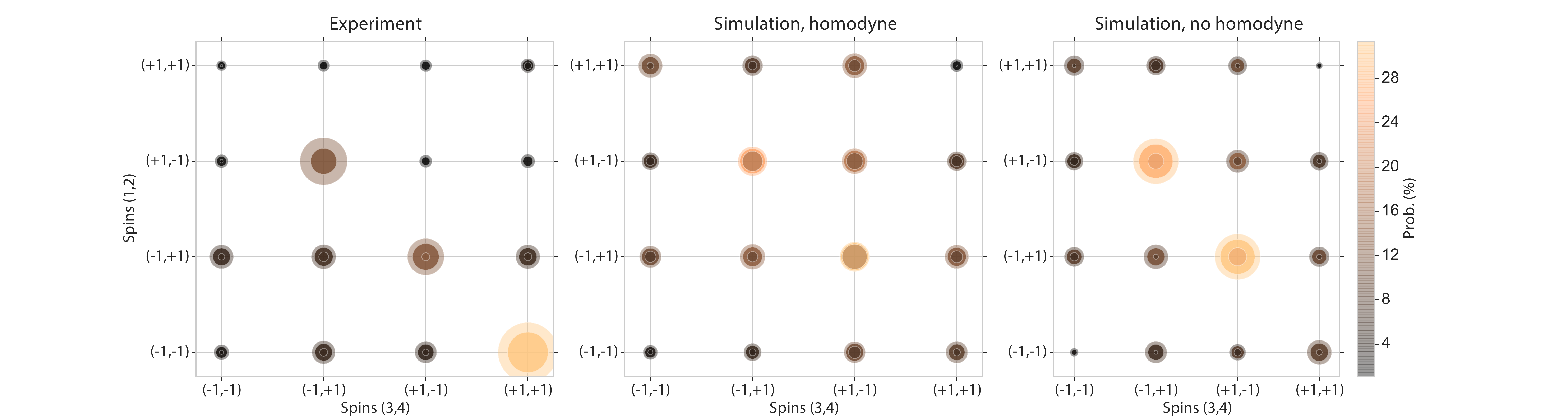}
        \caption{$\alpha = 1$, antiferromagnet}
        \label{fig:antiferro_DD1}
    \end{subfigure}
        \begin{subfigure}[t]{\textwidth}
        \includegraphics[width=\textwidth]{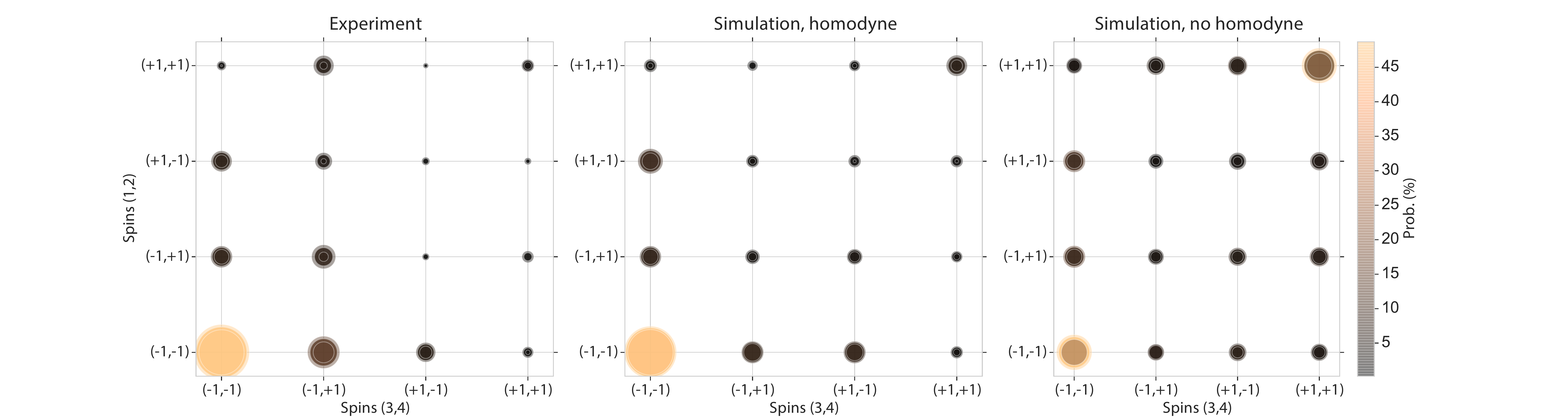}
        \caption{$\alpha = 0$, ferromagnet}
        \label{fig:ferro_DD0}
    \end{subfigure}
    \begin{subfigure}[t]{\textwidth}
        \includegraphics[width=\textwidth]{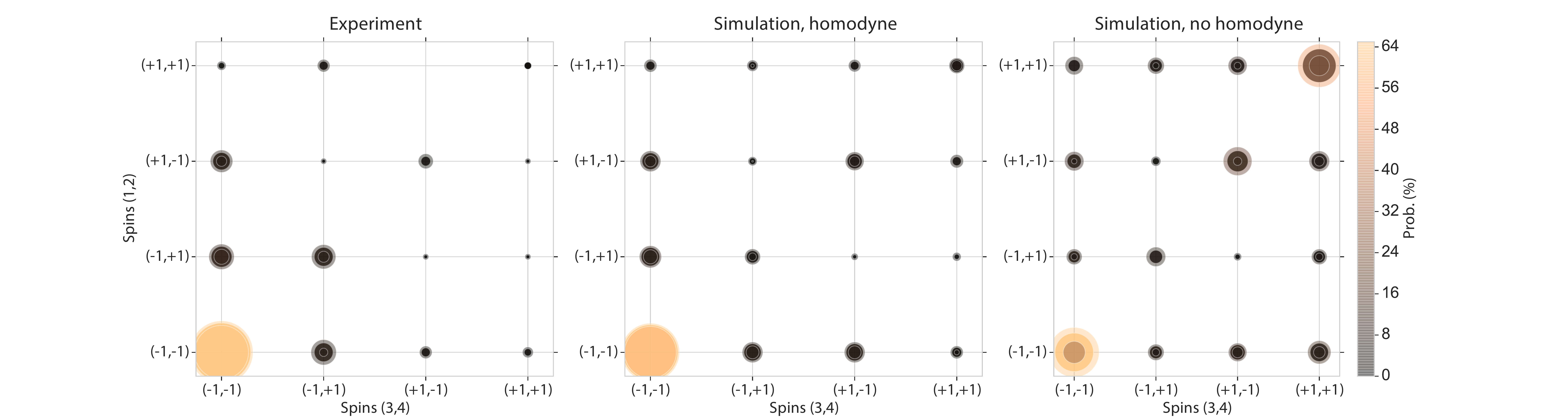}
        \caption{$\alpha = 1$, ferromagnet}
        \label{fig:ferro_DD1}
    \end{subfigure}
    \caption{Phase space representation of antiferromagnet with PNP (Left), simulation with homodyne detection (center), and simulation without homodyne detection (right).}
    \label{fig:antiferro}
\end{figure}
\begin{figure}
    \centering
    \begin{subfigure}[t]{\textwidth}
        \includegraphics[width=\textwidth]{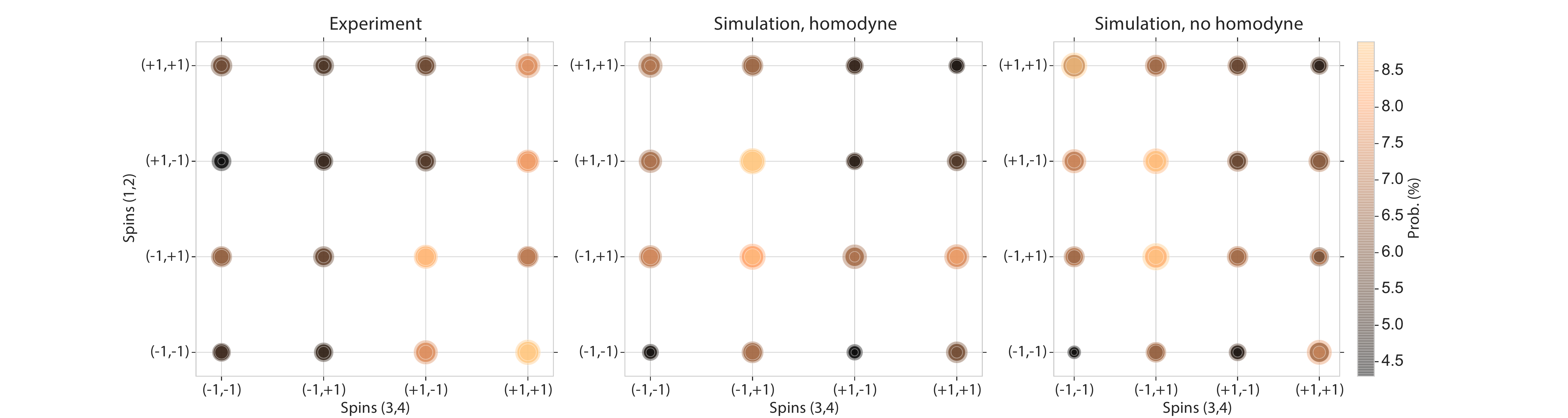}
        \caption{$\alpha = 0$, random cubic graph}
        \label{fig:randomcubic_DD0}
    \end{subfigure}
    \begin{subfigure}[t]{\textwidth}
        \includegraphics[width=\textwidth]{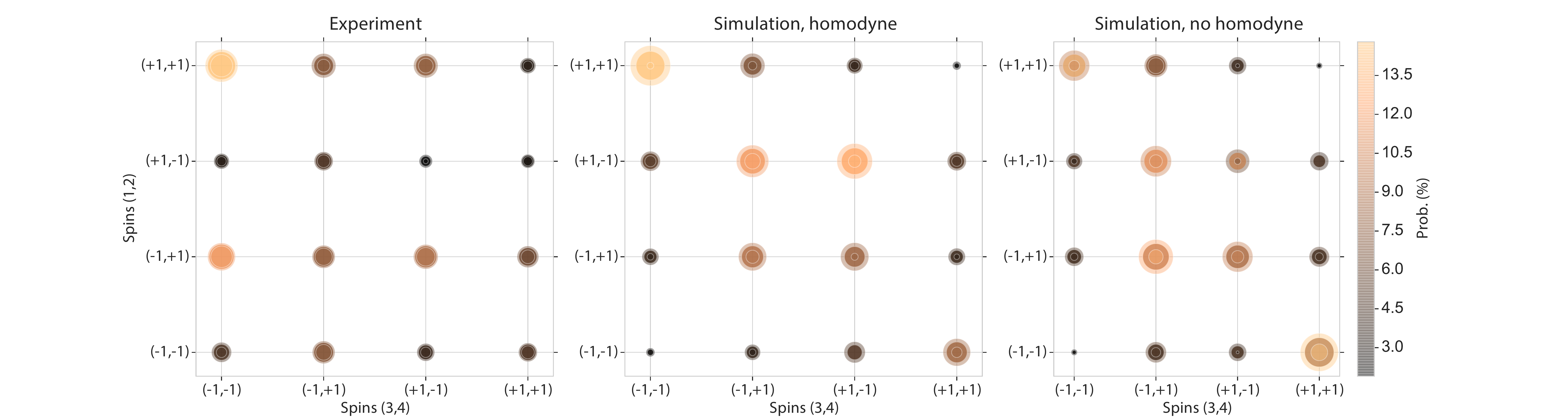}
        \caption{$\alpha = 1$, random cubic graph}
        \label{fig:randomcubic_DD1}
    \end{subfigure}
    \begin{subfigure}[t]{\textwidth}
        \includegraphics[width=\textwidth]{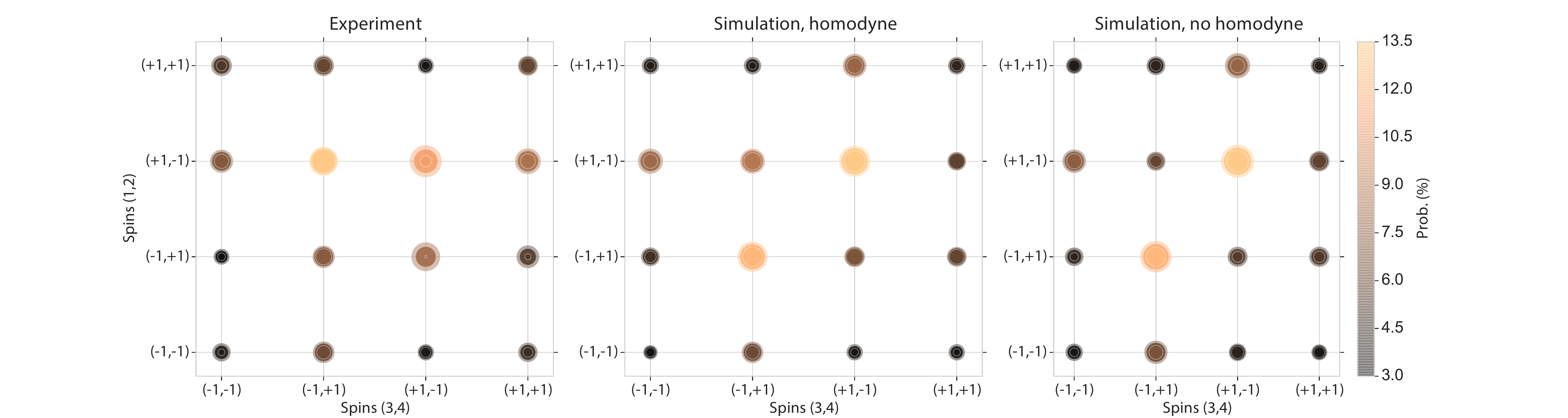}
        \caption{$\alpha = 0$, random sparse graph}
        \label{fig:randomsparse_DD0}
    \end{subfigure}
    \begin{subfigure}[t]{\textwidth}
        \includegraphics[width=\textwidth]{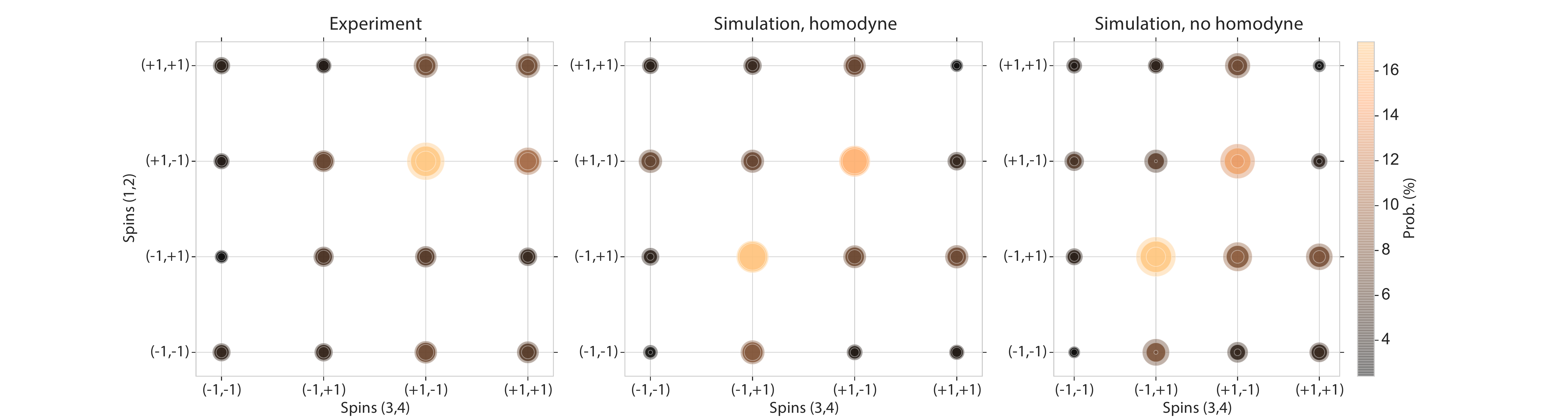}
        \caption{$\alpha = 1$, random sparse graph}
        \label{fig:randomsparse_DD1}
    \end{subfigure}    
    \caption{Phase space representation of random cubic graph with PNP (Left), simulation with homodyne detection (center), and simulation without homodyne detection (right).}
    \label{fig:ferro}
\end{figure}
\begin{figure}
    \centering
    \begin{subfigure}[t]{\textwidth}
        \includegraphics[width=\textwidth]{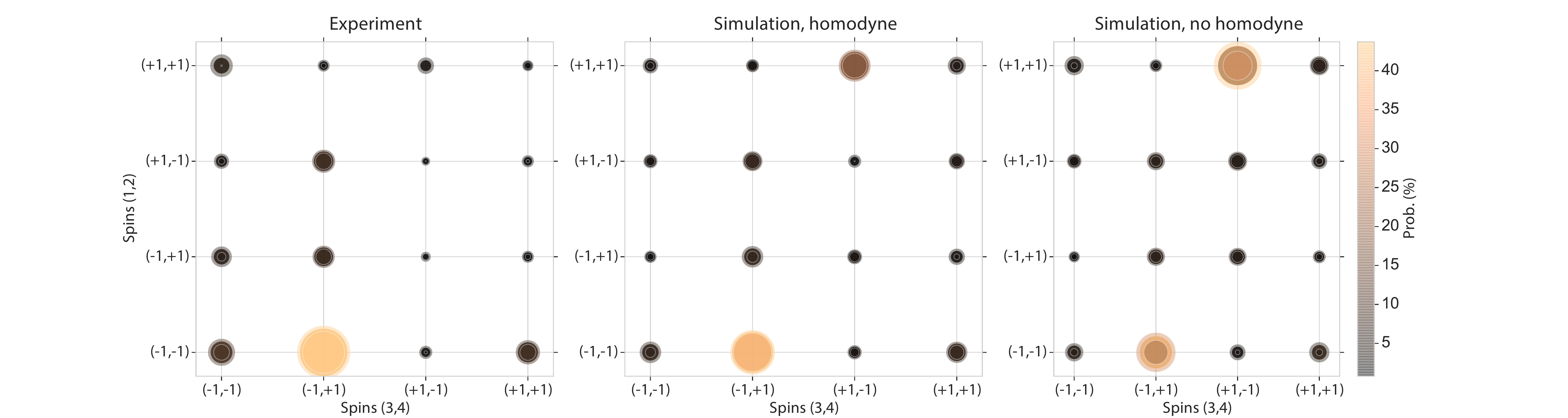}
        \caption{$\alpha = 0$, spin glass A}
        \label{fig:spinglass5_DD0}
    \end{subfigure}
    \begin{subfigure}[t]{\textwidth}
        \includegraphics[width=\textwidth]{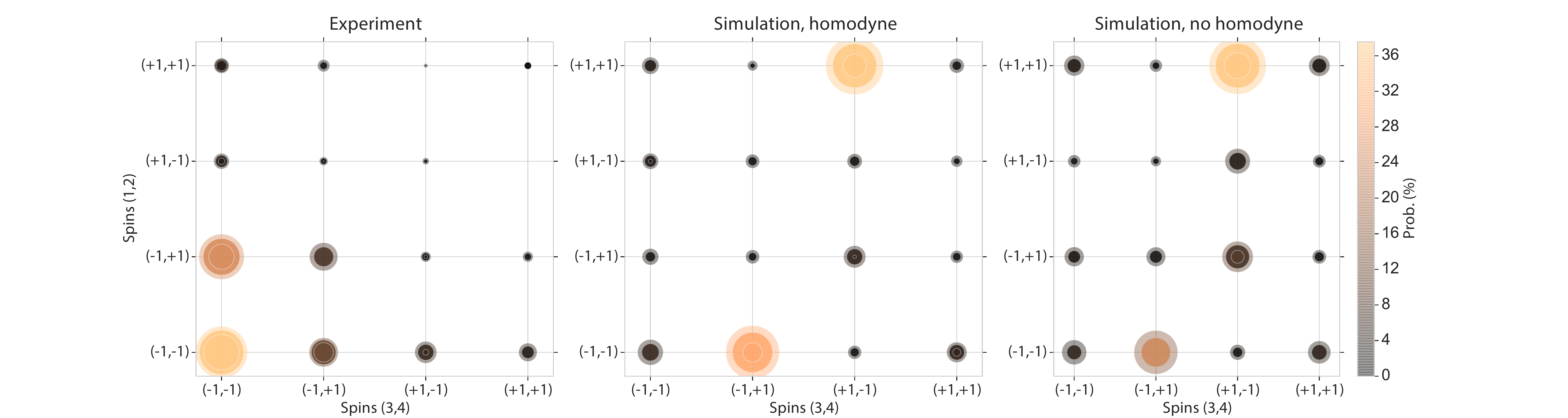}
        \caption{$\alpha = 1$, spin glass A}
        \label{fig:spinglass5_DD1}
    \end{subfigure}
    \begin{subfigure}[t]{\textwidth}
        \includegraphics[width=\textwidth]{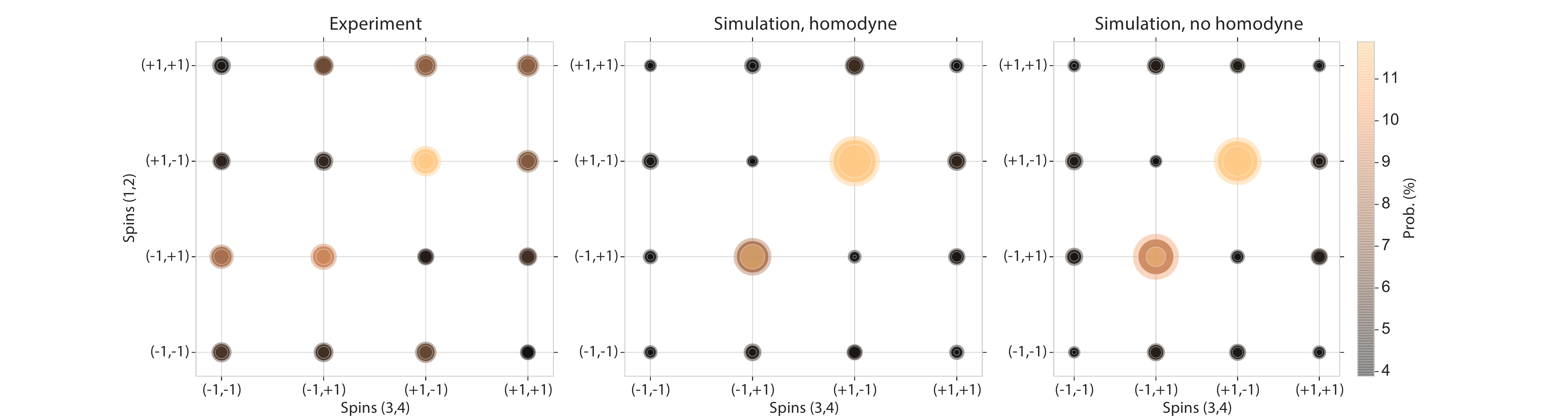}
        \caption{$\alpha = 0$, spin glass B}
        \label{fig:spinglass4_DD0}
    \end{subfigure}
    \begin{subfigure}[t]{\textwidth}
        \includegraphics[width=\textwidth]{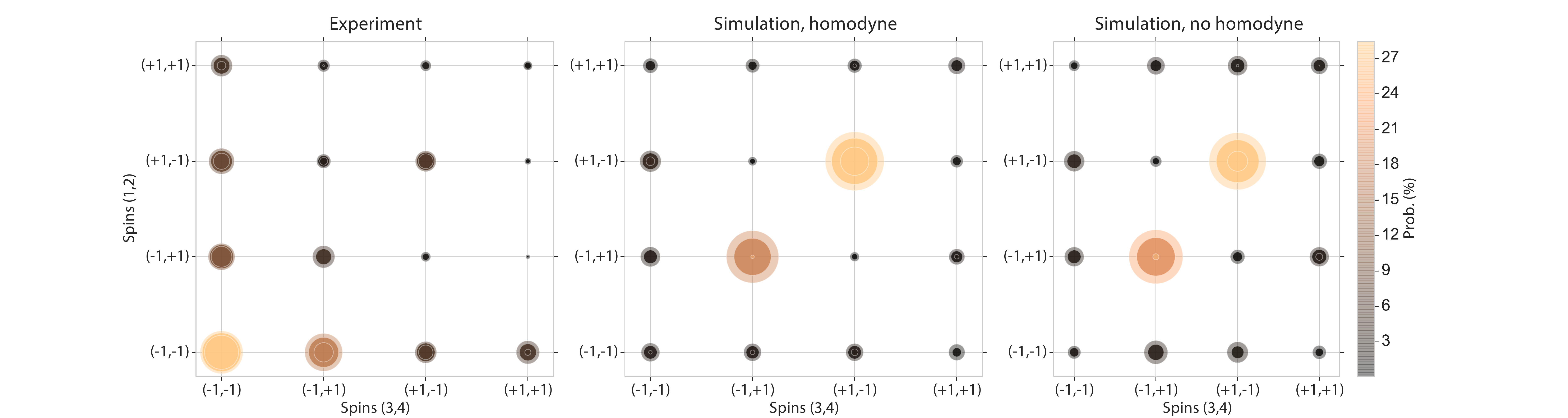}
        \caption{$\alpha = 1$, spin glass B}
        \label{fig:spinglass4_DD1}
    \end{subfigure}    
    \caption{Phase space representation of spin glass A with PNP (Left), simulation with homodyne detection (center), and simulation without homodyne detection (right).}
    \label{fig:ferro}
\end{figure}
\begin{figure}
    \centering
    \begin{subfigure}[t]{0.4\textwidth}
    \centering
        \includegraphics[width=1.1\textwidth]{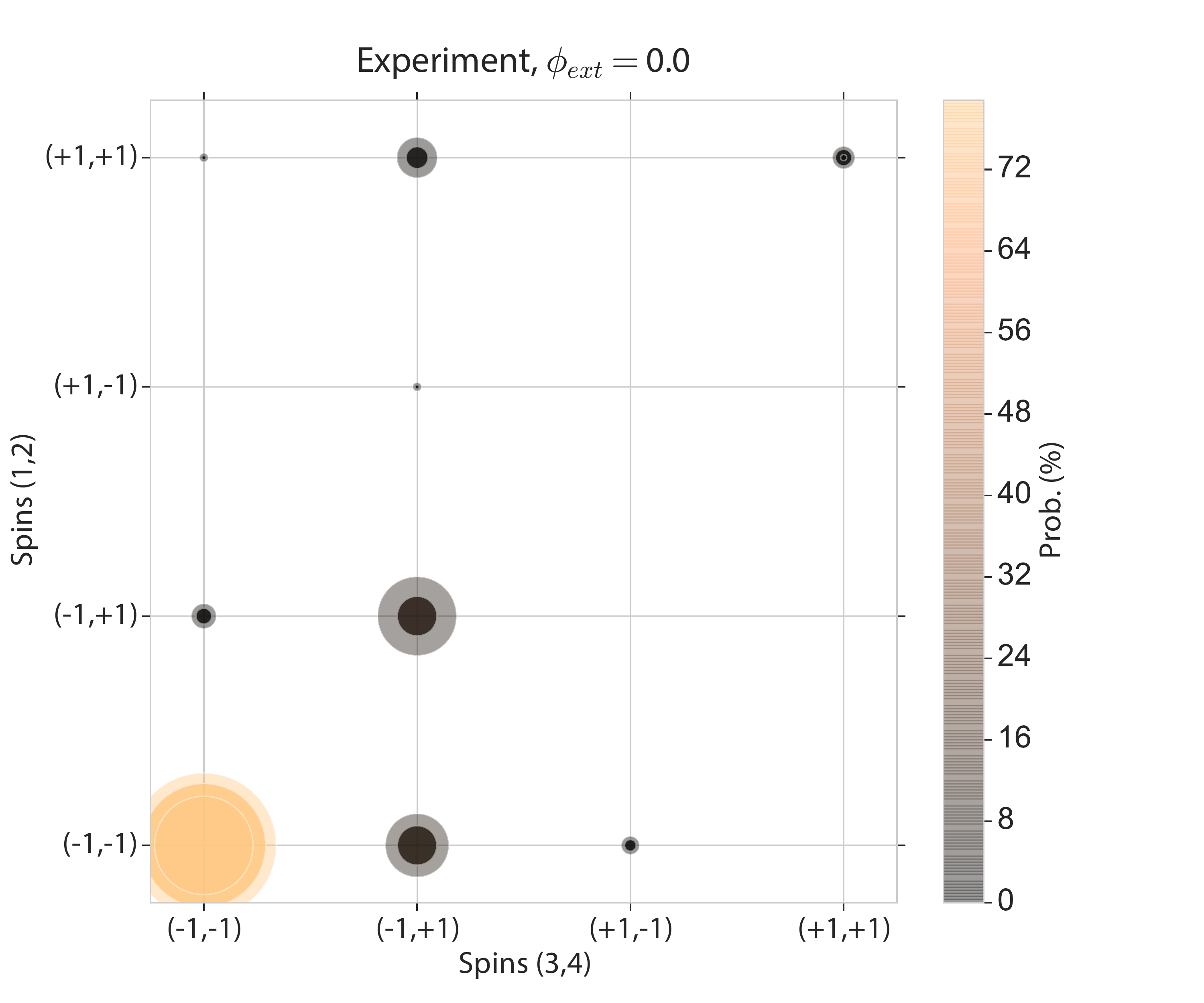}
        \label{fig:0pt0phase_space_expt}
    \end{subfigure}
    \begin{subfigure}[t]{0.4\textwidth}
    \centering    
        \includegraphics[width=1.1\textwidth]{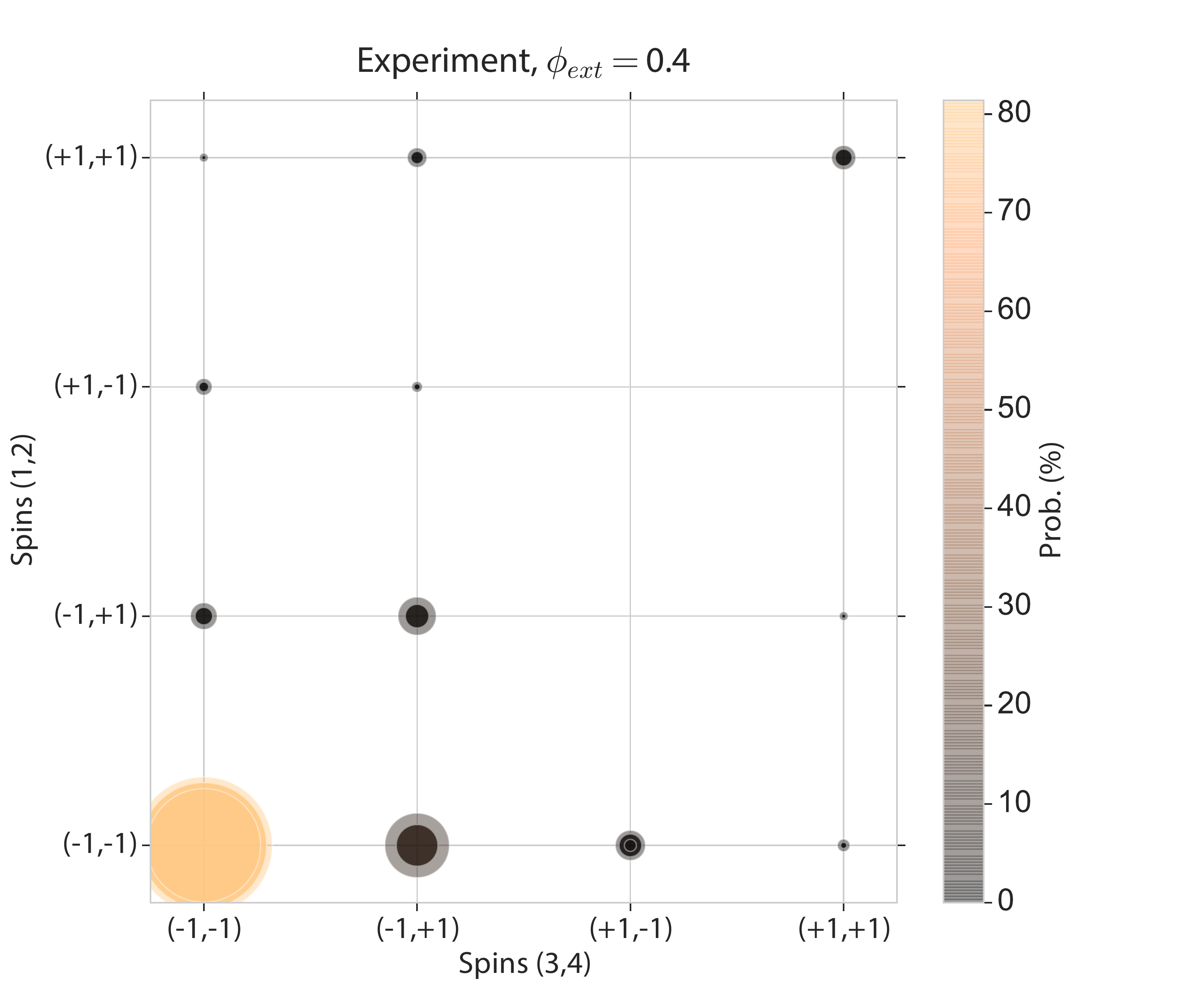}
        \label{fig:0pt4phase_space_expt}
    \end{subfigure}       
    \begin{subfigure}[t]{0.4\textwidth}
    \centering    
        \includegraphics[width=1.1\textwidth]{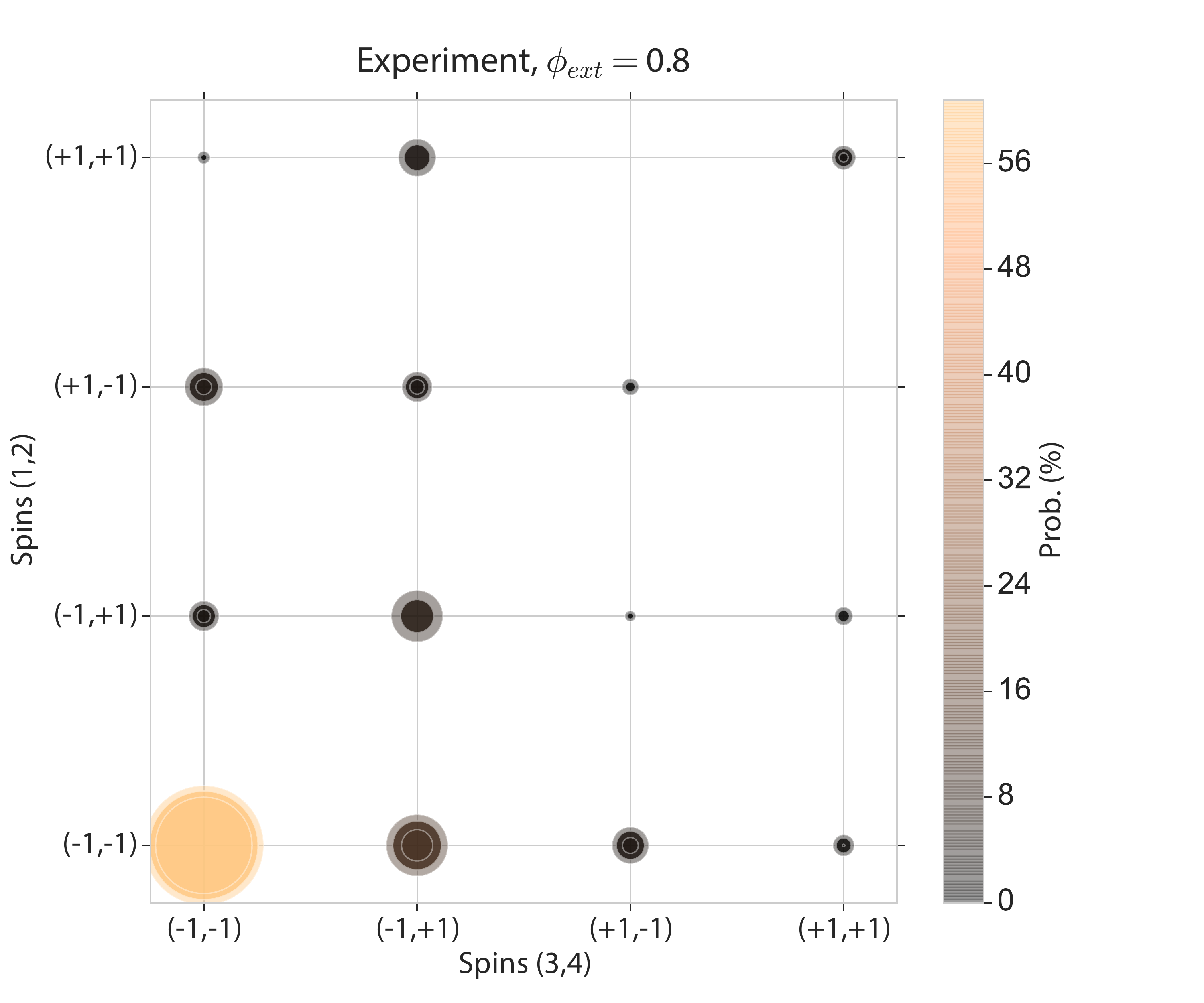}
        \label{fig:0pt8phase_space_expt}
    \end{subfigure}    
    \begin{subfigure}[t]{0.4\textwidth}
    \centering    
        \includegraphics[width=1.1\textwidth]{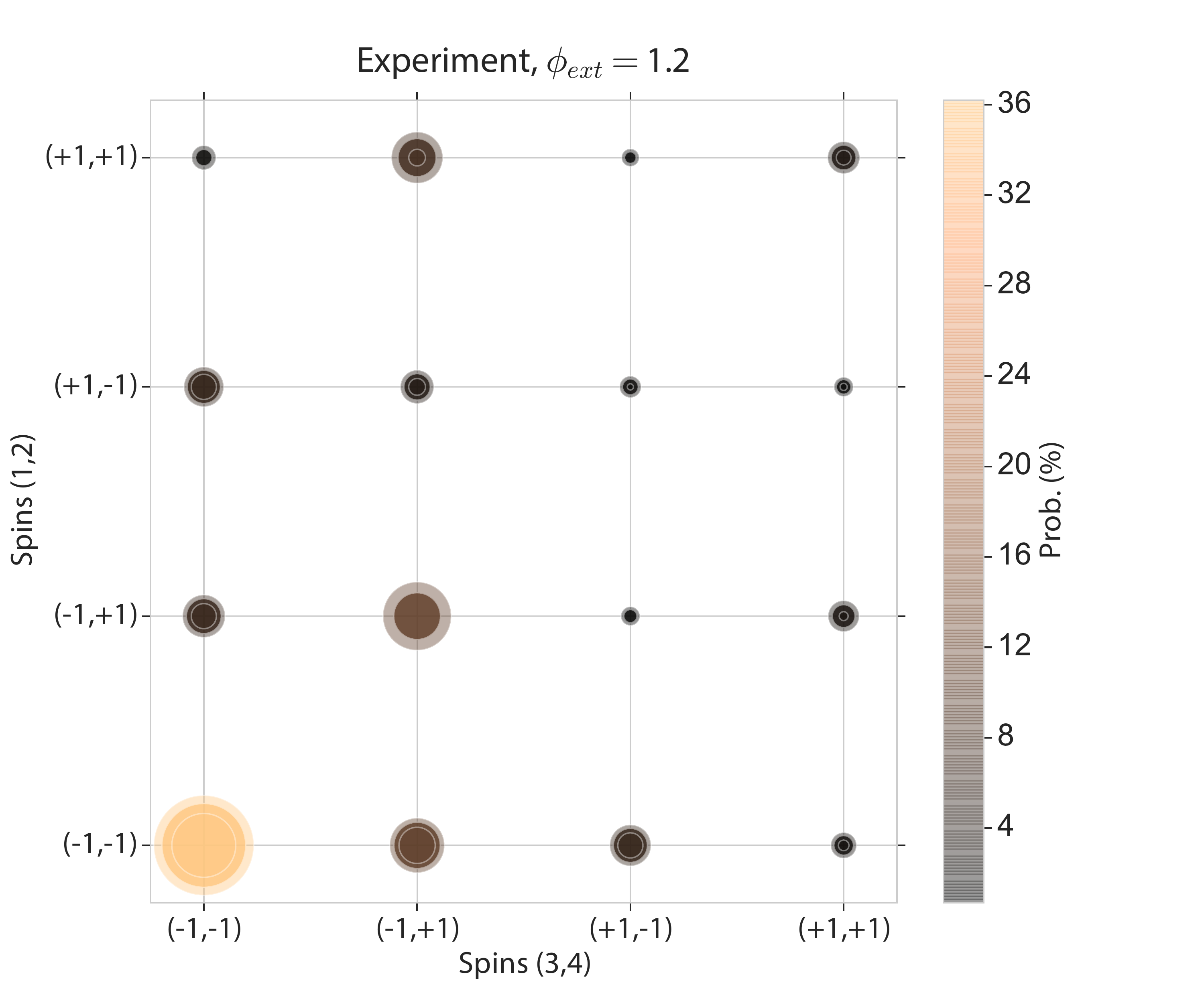}
        \label{fig:1pt2phase_space_expt}
    \end{subfigure}               
    \begin{subfigure}[t]{0.4\textwidth}
    \centering    
        \includegraphics[width=1.1\textwidth]{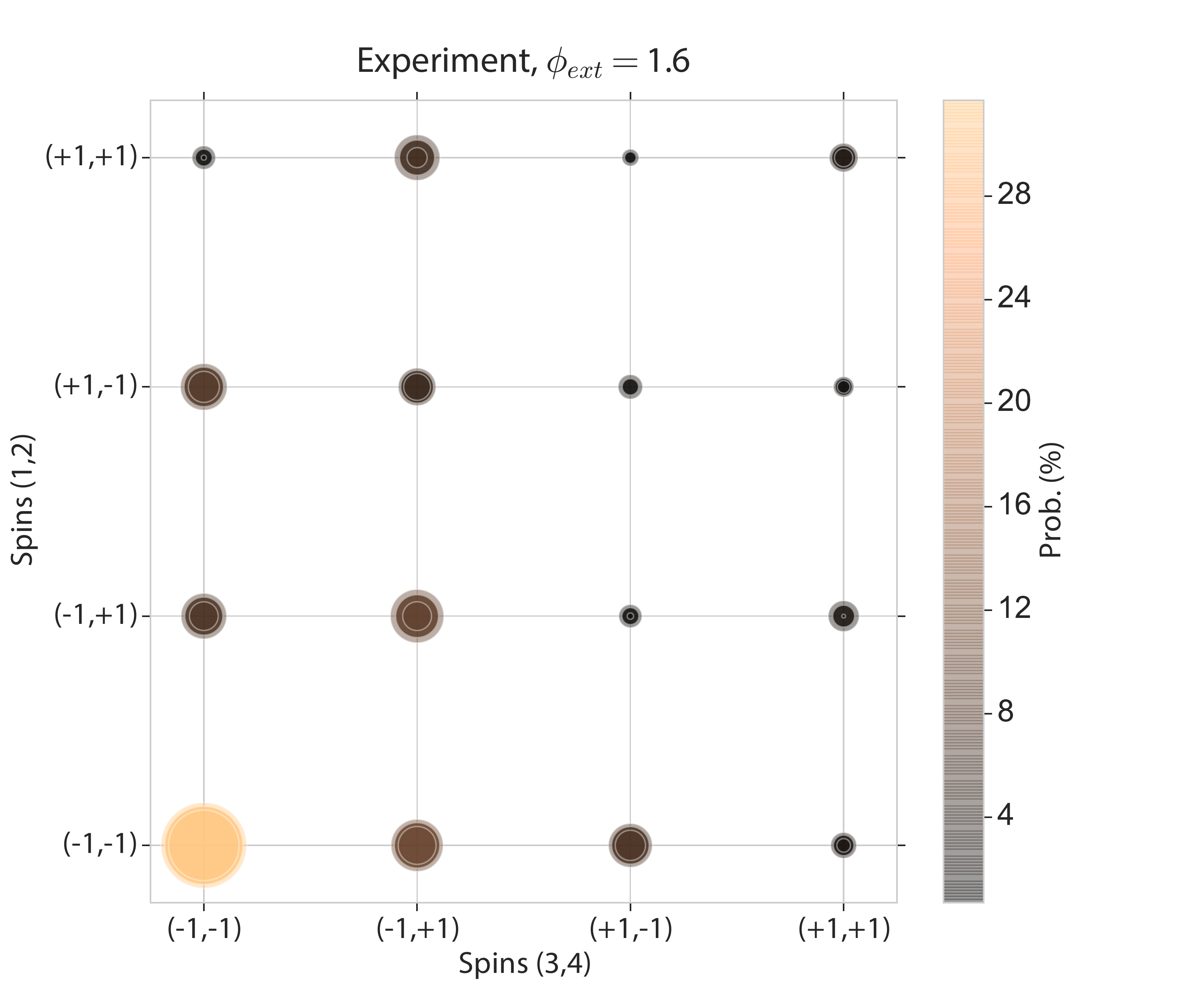}
        \label{fig:1pt6phase_space_expt}
    \end{subfigure}          
    \begin{subfigure}[t]{0.4\textwidth}
    \centering    
        \includegraphics[width=1.1\textwidth]{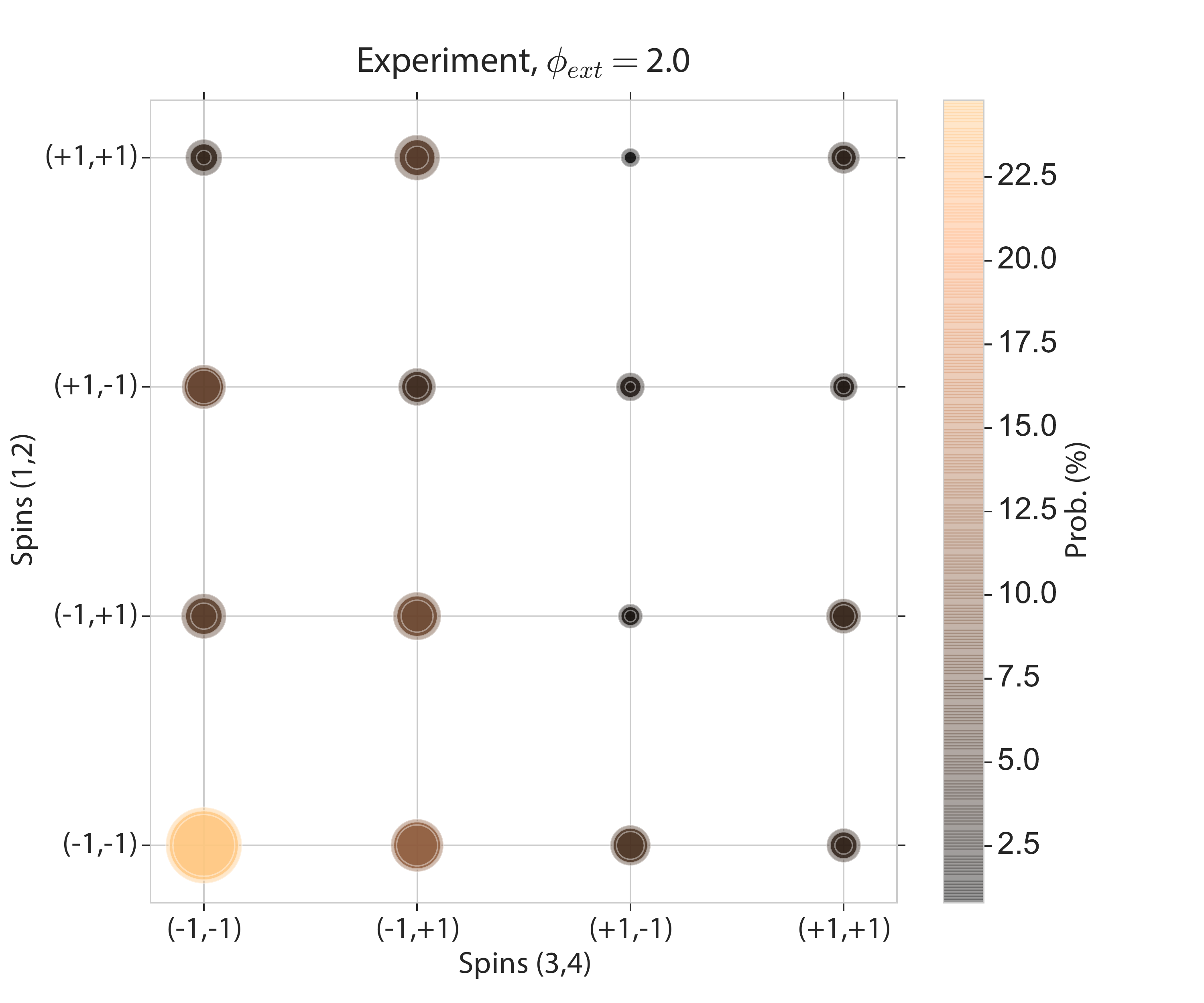}
        \label{fig:2pt0phase_space_expt}
    \end{subfigure}                                   
    \caption{Phase space exploration for ferromagnet and various noise levels.}
    \label{fig:antiferro}
\end{figure}

\clearpage
\bibliographystyle{unsrt}
\bibliography{supplement}